\documentclass{emulateapj}
\def\stacksymbols #1#2#3#4{\def\theguybelow{#2}
        \def\verticalposition{\lower#3pt}
        \def\spacingwithinsymbol{\baselineskip0pt\lineskip#4pt}
        \mathrel{\mathpalette\intermediary#1}}
\def\intermediary #1#2{\verticalposition\vbox{\spacingwithinsymbol
        \everycr={}\tabskip0pt
        \halign{$\mathsurround0pt#1\hfil##\hfil$\crcr#2\crcr
                \theguybelow\crcr}}}
\def\lta{\stacksymbols{<}{\sim}{2.5}{.2}}
\def\gta{\stacksymbols{>}{\sim}{3}{.5}}

\shorttitle{EVOLUTION OF CYGNUS A}
\shortauthors{MATHEWS \& GUO}

\begin{document}

\title{SELF-CONSISTENT EVOLUTION OF GAS AND COSMIC RAYS 
IN CYGNUS A AND SIMILAR FR II CLASSIC DOUBLE RADIO SOURCES}

\author{William G. Mathews\altaffilmark{1} and 
Fulai Guo\altaffilmark{1}}

\altaffiltext{1}{University of California Observatories/Lick
Observatory,
Department of Astronomy and Astrophysics,
University of California, Santa Cruz, CA 95064 
mathews@ucolick.org}

\begin{abstract}
In Cygnus A and other classical FR II double radio sources, powerful
opposing jets from the cores of halo-centered galaxies drive out into
the surrounding cluster gas, forming hotspots of shocked and
compressed cluster gas at the jet extremities.  The moving hotspots
are sandwiched between two shocks.  An inner-facing shock receives
momentum and cosmic rays from the jet and creates additional cosmic
rays that form a radio lobe elongated along the jet axis.  An
outer-facing bow shock moves directly into the undisturbed group or
cluster gas, creating a cocoon of shocked gas enclosing the radio
lobe.  We describe computations that follow the self-consistent
dynamical evolution of the shocked cluster gas and the relativistic
synchrotron-emitting gas inside the lobes.  Relativistic and
non-relativistic components exchange momentum by interacting with
small magnetic fields having dynamically negligible energy densities.
The evolution of Cygnus A is governed almost entirely by cosmic ray
energy flowing from the hotspots. Mass flowing into hotspots from the
jets is assumed to be small, greatly reducing the mass of gas flowing
back along the jet, common in previous calculations, that would
disrupt the spatial segregation of synchrotron-loss ages observed
inside FR II radio lobes.  We compute the evolution of the cocoon when
the velocity and cosmic ray luminosity of the hotspots are constant
and when they vary with time.  If cosmic rays mix with cluster gas in
hotspots before flowing into the radio lobe, the thermal gas is heated
to mildly relativistic temperatures, producing an unobserved pressure
inside the lobe.
\end{abstract}

\vskip.1in

\section{Introduction}

Figure 1 shows a $\sim$200 ks Chandra image of
the archetypical twin-jet FR II\footnotemark[2] source Cygnus A
and a second X-ray image with radio contours at 5GHz superimposed 
(Wilson, Smith \& Young 2006).
The bright core of Cygnus A is coincident with the 
nucleus of a massive galaxy 
at the center of a cluster with
mass exceeding $\sim3 \times  10^{14}$ $M_{\odot}$ 
(Smith et al. 2002).
We adopt a distance $\sim230$ Mpc to Cygnus A 
at which $1^{\arcmin\arcmin}$ corresponds to 1 kpc.

As first discussed by Blandford \& Rees(1974) and Scheuer (1974),
Cygnus A and other FR II radio sources are formed by
two essentially identical oppositely-directed jets
created near cluster-centered black holes 
that penetrate the cluster gas at
mildly relativistic speeds $\approx 0.01 - 0.1c$.
The jets impact the cluster gas in strong (reverse) shocks
at the inner boundaries of bright kpc-sized ``hotspots'' visible
in Figure 1 at the tips of the
radio and X-ray images about 60 kpc from the center.
The supersonic motion of these relatively dense hotspots through the
ambient cluster gas forms symmetric outward-facing 
bow shocks that propagate directly into undisturbed cluster gas.
The football-shaped bow shock forms a cocoon of shocked cluster
gas that encloses the radio jets and lobes.
The age of Cygnus A, about $10^7$ yrs, can be 
estimated from the observed age of the oldest radio-synchrotron 
electrons (e.g. Machalski et al. 2007).
From this age and the hotspot-center distance, 
the average velocity of 
the hotspots is about $v_{hs}\sim6000$ km s$^{-1}$ 
which should also be consistent with the strength 
of the cocoon bow shock.
The total power being delivered to the hot gas is $\sim10^{46}$
erg s$^{-1}$, as suggested by Wilson et al. (2006).

\footnotetext[2]{Fanaroff-Riley extragalactic radio source types 
I and II are distinguished by their radio
morphology: FR I (FR II) being fainter (brighter)
in radio emission in the distant lobes than near the central jets
and AGN.}

We describe here approximate computations of the evolution 
of Cygnus A -- and FR II sources in general -- that include  
for the first time dynamical interactions between the hot cluster 
gas and the relativistic fluid confined inside the radio lobes.
In view of the unknown nature and contents of the Cygnus A jets 
and the very small volume they occupy, we regard 
the hotspot as the principal source of energy. 
As a hotspot moves at some assumed velocity away from the 
cluster center, gas inside the hotspot receives a local compression 
from the jet impact. 
Moreover, the jet-hotspot shock transmits cosmic rays from the jet 
and accelerates additional cosmic rays associated with the 
shock compression. 
By regarding the hotspots as sources of dynamic and relativistic 
energy we avoid a detailed computation of the flow of energy 
and relativistic particles along the jet about which rather little 
is known from observation.

\section{The Radio Lobes}

The relativistic electrons responsible for the non-thermal 
radio synchrotron emission in Figure 1 
are assumed to be transported to the hotspot 
in the jets or produced locally in the
strong jet-hotspot shock, or both. 
In a weakly magnetized plasma 
electrons and protons in the high energy tail of
the Maxwellian thermal distribution can be accelerated to relativistic
energies by multiple scatterings across strong shock waves 
(Blandford \& Eichler 1987).
Cosmic ray acceleration 
increases with the shock duration and Mach number.
For this reason we see 5 GHz synchrotron emission from
relativistic electrons streaming away from
powerful jet-driven shocks in the luminous hotspots
in Figure 1, but not from the much weaker
bow shocks (Mach $\sim 1.3 - 2.5$) that enclose the cocoon.
Evidently hotspots receive a huge momentum kick and compression
at their inner edges where the jet impact occurs.
As a result, gas flows away from the hotspots at a high, nearly sonic
velocity in all directions
roughly perpendicular to the jet axis, carrying gas and newly
accelerated relativistic electrons with it.
This powerful transverse flow contributes to the width of the
cocoon-enclosing shock perpendicular to the jet axis
and the width of the radio-emitting lobes.

The double-lobe radio image at 5 GHz in Figure 1 
does not show the full extent of the radio lobes in Cygnus A. 
Non-thermal mission at lower radio frequencies 
extends all the way to the center where the two 
lobes merge, 
maintaining an approximately uniform width 
(e.g. Steenbrugge et al. 2010).
While FR II sources like Cygnus A are often 
referred to as classical radio double sources, 
many authors use the term ``bridge'' 
if the low frequency radio emission extends 
continuously to the cluster center. 
The 5 GHz lobes in Figure 1 appear truncated at their 
inner boundaries because 
the electrons that radiate most efficiently 
at this (relatively high) frequency 
are no longer present closer to the 
center of Cygnus A, reflecting a loss of electron energy 
and possible variations in the magnetic field.
In fields of a few $\mu$G
electrons of energy $\gamma = E/m_e c^2 \sim 10^3-10^5$
emit radio synchrotron radiation
concentrated near frequency $\nu_c = \gamma^2 B (e/2 \pi m_e c)$
and have lifetimes $t_{age} \approx (3 m_e c / 4\sigma_T)/\gamma u$
where $u = (B^2 + B_{cmb}^2)/8\pi$ is the energy density 
and $B_{cmb} = 3\mu$G is the
equivalent magnetic field for CMB inverse Compton (IC-CMB) losses.
The entire region between opposing hotspots is filled
with radio emission having progressively steeper
radio spectra near the center where older electrons, 
ejected from the hotspot at earlier times, radiate at lower $\nu$.
This radio-wake provides essential information about the 
past velocity of the reverse shock and hotspot and the
local magnetic field affecting $t_{age}$.

Placing a transparent ruler on the Chandra image in Figure 1
reveals that (1) the brightest hotspots on E and W sides
are exactly collinear with the central galactic source
and (2) the smaller hotspot in the W is aligned with
faint, broad X-ray ``jets'' on both sides.
Furthermore, the current direction of the very thin one-sided western
(Doppler-boosted approaching) jet, 
visible in the radio image, is terminated by
none of the hotspots visible on the W side.
Evidently the jet direction varies with time.

Observations of inverse Compton (IC-CMB) X-ray emission 
arising from interactions of radio-synchrotron electrons 
with  the cosmic microwave
background, when combined with low frequency radio synchrotron 
observations from cosmic ray electrons having the same energy, 
allow a direct determination of the magnetic field strength. 
Moreover, if the radio lobes in Cygnus A are assumed to be in 
approximate pressure equilibrium with 
their hot cocoon gas environment of known or estimated pressure, 
it is possible to estimate the energy density of the 
relativistic particles in the radio lobes.
By this means it is found that the 
relativistic electron energy density greatly exceeds 
the magnetic field energy density
by factors of 10-600 (Hardcastle \& Croston 2010; Yaji et al. 2010).
It is remarkable that 
the X-ray image in Figure 1 shows no clear evidence of a cavity
associated with the 5GHz radio emission, 
although the crowding of radio contours at the lobe boundaries
indicates a sharp contact discontinuity between 
cosmic rays and the thermal gas.
Evidently the X-ray cavity is filled in with IC-CMB (and possibly 
also synchrotron self-Compton) emission 
that fortuitously matches the thermal X-ray emission just outside 
the lobes so there is no easily visible X-ray cavity.
Normally, one expects some diffusion of relativistic 
cosmic ray electrons, 
but the sharp radio lobe boundary indicates that 
very little diffusion occurs at this interface with the thermal gas.
In addition, 
the stratification of radio-synchrotron electrons of different ages
and energies along the Cygnus A radio cavity of age $10^7$ yrs,
implies an upper limit on the CR diffusion coefficient 
inside the radio cavity, 
$\kappa < (60~{\rm kpc})^2/(10^7~{\rm yrs}) \approx 10^{32}$
cm$^2$ s$^{-1}$, which seems easy to satisfy.

\section{The Phantom Hotspot}

The evolution of FR II radio sources similar to those in Cygnus A 
has been the subject of many detailed computations 
(e.g. Carvalho et al. 2005 
or O'Neill \& Jones 2010 and references therein).
Nevertheless, there is little consensus about 
the internal nature or contents of the jets -- 
electron pairs, cosmic rays, magnetic field, thermal gas, etc.
Perhaps the jets are initially purely electromagnetic but 
entrain some cluster gas and inertia as they progress outward.
In addition, internal shocks perhaps 
arising from perturbations or 
changes in the shape of the jet wall geometry, 
can rejuvenate and accelerate new relativistic particles inside jets.
Nevertheless, for computational expediency 
in most or all previous computations of FR II jets 
the jet content is assumed to be thermal gas, sometimes   
with adiabatic index $\gamma = 4/3$ rather than $5/3$. 
However, a significant mass of thermal gas flowing out in the jet
invariably results in fast ``backflows'', 
a contrary flow just outside the jet boundary that flows 
anti-parallel to the outgoing jet.
After passing through the reverse shock in the hotspot, 
mass-carrying jets encounter 
high-pressure cluster gas locally compressed near the moving hotspot. 
This produces a positive pressure gradient in the 
gas just outside the jet (opposite to the negative 
pressure gradient in the undisturbed cluster gas) 
that drives backflows of shocked gas 
back toward the center of the cluster.
Such computed backflows can 
generate strong Kelvin-Helmholtz (KH) instabilities and general
turmoil which is not observed in FR II radio cocoons. 

Thermal gas backflows are described in some detail by Krause (2005)
who considered so-called ``light'' jets with internal densities that are
$\sim 10^{-4}$ below that of the initial central cluster gas.
The (negative) backflow velocity is fast, typically exceeding  
in magnitude the outward velocity of the hotspot 
-- by nearly ten in the FR II calculations of Reynolds et al. (2002).
However, strong backflows appear to be inconsistent with 
radio observations. 
In particular, computed backflowing gas 
moving through the radio cavity region becomes highly disordered
by shear-driven KH instabilities. 
Such backflows would advect and upset the radially ordered 
age-related energy distributions of 
radio-synchrotron electrons observed along the radio lobes 
by Carilli et al (1991) 
and illustrated by Steenbrugge et al. (2010).
Moreover, the sharply defined outer boundaries
of the Cygnus A radio lobes in Figure 1 would not 
in general be possible.
Finally, the faint irregular X-ray jet-like features 
visible in Figure 1, 
whatever their uncertain origin, and also the very narrow and 
sharply defined radio jet appear to be completely undisturbed
by turbulent backflow activity similar to that predicted  
by most, possibly all, previous FR II calculations.

To avoid this undesirable outcome, 
we consider here jets that carry a very small amount of 
non-relativistic gas and which occupy such a small volume, 
as in the radio image in Figure 1, 
that they can be ignored altogether, 
i.e. our computation is driven not by the jet but by  
hotspots energized by the jet.
Although our understanding of the physical environment 
within hotspots is still very incomplete, 
because of their relatively high luminosity and resolvable structure 
more is known about the internal physics of hotspots 
(e.g. Stawarz et al. 2007) 
than in the jets themselves.

The FR II hotspot energy source in our computations
is represented with a ``phantom'' hotspot compression 
that moves out in the computational grid in the jet direction 
and is also the source of cosmic ray energy.
When the hotspot compression moves supersonically relative 
to the cluster gas,
bow shocks form, producing cocoons of shocked gas 
as observed in Cygnus A.
While there is no compelling observational 
evidence of deceleration in FR II sources 
( O'Dea et al. 2009) 
we nevertheless entertain this possibility.
The phantom hotspot velocity is parameterized as follows:
\begin{equation}
v_{hs} = (v_0 - v_{\infty})e^{-\tau/\tau_0} + v_{\infty}
\end{equation}
where 
\begin{equation}
\tau = t/t_a 
\end{equation}
is the time normalized by 
the current age $t_a$ of the FR II structure 
and  $v_0$ is the initial velocity.
The time parameter $t_0 = \tau_0 t_a$ defines 
the deceleration epoch of the hotspot compression
which moves with uniform velocity $v_0$ 
if $\tau_0 = t_0/t_a$ is assumed to be very large.
The average velocity at any time is 
\begin{equation}
\langle v_{hs} \rangle = {\tau_0 \over \tau}(v_0 - v_{\infty})
(1 - e^{-\tau/\tau_0}) + v_{\infty}
\end{equation}
and $r_{hs} = \langle v_{hs} \rangle t$ 
is the radial position of the hotspot at any time.
The final velocity $v_{\infty}$ is not a free parameter 
but instead ensures that the mean velocity is 
$\langle v_{hs} \rangle_a = r_a/t_a$ at time $t_a$ for any choice of 
$v_0$ and $t_0$, i.e.
\begin{displaymath}
v_{\infty} = { \langle v_{hs} \rangle_a - v_0 \tau_0 (1-e^{1/\tau_0})
\over 1 - \tau_0 (1-e^{1/\tau_0}) }.
\end{displaymath}
We assume that the volume of the phantom hotspot $V_{hs}$ 
remains constant during its evolution.
When the cosmic ray power density ${\dot S}_{cr}$
injected into the moving hotspot is 
uniform in space and time, then the rate of change of
the energy density of cosmic rays in the hotspot $e_{c,hs}$ is
also constant 
\begin{equation}
{\dot S}_{hs} = {de_{c,hs} \over dt} = {L_{cr} \over V_{hs}}
~~~{\rm erg~cm}^{-3}{\rm ~s}^{-1}
\end{equation}
where $L_{cr}$ is the mean cosmic ray luminosity of the hotspot
over time $t_a$. 
However, when the hotspot has a non-uniform velocity, 
the phantom hotspot receives the same total energy 
in cosmic rays over time $t_a$, but at a variable rate,
\begin{equation}
{\dot S}_{hs} = {\langle v_{hs} \rangle_a \over v_{hs}(\tau)} 
{L_{cr}\over V_{hs}}~~~{\rm erg~cm}^{-3}{\rm ~s}^{-1}.
\end{equation}

While the calculations we describe 
here are not intended to exactly reproduce 
the observed radio and X-ray properties of Cygnus A, 
we choose parameters that approximate those of 
this relatively nearby and well-observed FR II source.
The current age of Cygnus A, $t_a = 10^7$ yrs, is taken from a recent
detailed analysis of the radio spectrum distribution 
along the radio lobe axes 
(Machalski et al. 2007).
The current distance of the Cygnus A hotspot from the 
galactic core, $r_a \approx 60$ kpc,
is based on the assumption that the source is oriented in the 
plane of the sky although 
the higher visibility of radio jets on the western side 
suggests that this side may be aligned toward us.
With these parameters the mean velocity of the hotspot,  
$\langle v_{hs} \rangle_a = r_a/t_a = 5870$ km s$^{-1}$, 
is well in excess of the sound speed in the hot cluster gas,
$c_s \approx 1100 (T/{4.6 \rm{ keV}})^{1/2}$ km s$^{-1}$.

Also of interest is the possibility that the cosmic ray energy 
provided to the Cygnus A hotspots by the jets and 
reverse shock can vary with time 
in a manner that is independent of the hotspot velocity.
FR II observational data suggest that 
neither the jet power nor the hotspot velocity varies with time
(O'Dea et al. 2009), but these statistical arguments 
are based on a sample of 31 FR II sources that contains  
very little information about hotspots at small 
distances, $\lta 50$ kpc 
that are most relevant to the past history of Cygnus A 
of current size $r_a \approx 60$ kpc. 
The size of FR II hotspots is observed to 
increase with the distance of the hotspot from the core 
of the host galaxies 
(Jeyakumar \& Saikia 2000; Perucho \& Marti 2003;
Kawakatu et al. 2008), 
where it is assumed that FR II sources evolve from 
the class of compact and medium-sized 
symmetric radio sources. 

If the cosmic ray power density received by the hotspot 
has an intrinsic time variation, this can be expressed 
with a properly normalized 
dimensionless function $\sigma_{hs}(\tau)$ of $\tau = t/t_a$,
\begin{equation}
{\dot S}_{hs} = 
{de_{c,hs} \over dt} = {L_{cr} \over V_{hs}}\cdot\sigma_{hs}(\tau)~~~
{\rm erg~cm}^{-3}{\rm ~s}^{-1}.
\end{equation}
The dimensionless time variation is assumed to be
\begin{equation}
\sigma_{hs}(\tau) ={  (1+\eta) + (1-\eta)\tanh[(\tau-\tau_e)/\Delta \tau]
\over (1+\eta) + (1-\eta) \Delta \tau 
\ln \left( {\cosh[(1-\tau_e)/\Delta \tau] 
\over \cosh[-\tau_e /\Delta \tau]}\right)   }
\end{equation}
where $\eta$, $\Delta \tau$ and $\tau_e$ are adjustable parameters.
If $\eta = 1$ then $\sigma_{hs} = 1$ and 
the jet power of the hotspot does not vary;  
if $\eta < 1$, the numerator in the expression for $\sigma_{hs}$ 
increases smoothly from 
$(1+\eta) + (1-\eta)\tanh[-\tau_e/\Delta \tau]$ to
$(1+\eta) + (1-\eta)\tanh[(1-\tau_e)/\Delta \tau]$ 
near time $\tau_e t_a$ 
during a time interval parameterized with $\Delta \tau$.
The denominator in equation (6) 
(which does not vary with $\tau$)
normalizes $\sigma_{hs}(\tau)$ so that 
\begin{equation}
\int_0^1 \sigma_{hs}(\tau) d\tau = 1,
\end{equation}
ensuring that the mean rate of cosmic ray injection into the hotspot 
at time $t_a$ is $L_{cr}/V_{hs}$ for any choice of parameters 
$\eta$, $\Delta \tau$ and $\tau_e$.

In \S5 below we describe how the gas and cosmic ray 
dynamics of FR II sources 
like Cygnus A can be completely determined by the assumed hotspot 
evolution in a given cluster environment.

\section{Computational Procedure}

\subsection{Equations of Cosmic Ray and Gas Dynamics}

Magnetic fields of strength 0.3-10 $\mu$G are ubiquitous
in cluster gas (Govoni \& Feretti 2004) and are generally stronger in
the enhanced feedback environment of cool-core clusters 
(Feretti et al. 2009).
The origin of these fields is controversial but only their
presence concerns us here.
Fields at this level, with energy densities
$B^2/8\pi = 6\times 10^{-13} (B/4\mu {\rm G})^2$ erg cm$^{-3}$
cannot significantly alter the dynamics of cluster gas with
thermal energy density
$3P/2 = 5 \times 10^{-11}(n_e/0.01~{\rm cm}^{-2})                                   
(T/{\rm keV})$ erg cm$^{-3}$.
Since $B^2$ and $P$ both decrease with cluster radius in about
the same way, $B^2/8\pi << 3P/2$ holds at every radius.
Magnetic fields are almost always dynamically subordinate 
in the cluster gas.

Cosmic rays are spatially confined by their
relativistic Larmor radii and 
are assumed to behave globally like a gaseous fluid.
Typical radio synchrotron cosmic ray electrons with energies
$\gamma \sim 10^3 - 10^5$ have very small gyroradii,
$r_g = \gamma mc^2/eB                                                               
\approx 4 \times 10^{12} (\gamma/10^4) (B/4\mu {\rm G})^{-1}$ cm.
Pressure gradients in the cosmic
ray fluid communicate momentum to fields and also to
the cluster gas since fields and gas are frozen together.
Most importantly,
cosmic rays and thermal gas can exchange momentum
even when the energy density of the communicating
field is very small, having no dynamical consequence of
its own.

The dynamical interaction of cosmic rays (CR) with
hot gas in groups/clusters can be
studied by solving the following equations: 
\begin{equation}
{ \partial \rho \over \partial t}
+ {\bf \nabla}\cdot\rho{\bf u} = 0
\end{equation}
\begin{equation}
\rho \left( { \partial {\bf u} \over \partial t}
+ ({\bf u \cdot \nabla}){\bf u}\right) =
- {\bf \nabla}(P + P_c) - \rho {\bf g} + \rho {\bf a}_{hs}
\end{equation}
\begin{equation}
{\partial e \over \partial t}
+ {\bf \nabla \cdot u}e = - P({\bf \nabla\cdot u})
\end{equation}
\begin{equation}
{\partial e_c \over \partial t}
+ {\bf \nabla \cdot u}e_c = - P_c({\bf \nabla\cdot u})
+ {\bf \nabla\cdot}(\kappa{\bf \nabla}e_c)
+ {\dot S}_{hs}
\end{equation}
where artificial viscosity terms are suppressed.
Pressures and thermal energy densities in the plasma
and cosmic rays are related respectively by
$P = (\gamma -1)e$ and $P_c = (\gamma_c - 1)e_c$.
The dynamics of the CRs are described by
$e_c$, the integrated energy density over the CR
energy or momentum distribution,
$e_c \propto \int E N(E) dE \propto \int p^4 f(p) (1+p^2)^{-1/2} dp$.
If desired, $e_c$ can refer to the relativistic CR energy
density for any combination of CR electrons or protons; 
we do not make this distinction in our models 
of Cygnus A, but synchrotron and IC 
radiation from relativistic electrons are obviously required.
Because of their negligible collective rest mass,
a mass conservation equation for the CRs is unnecessary.
Although CRs and hot plasma are coupled by mutual
interactions with a small magnetic field,
no magnetic terms need explicitly appear
in the equations provided magnetic stresses are small,
$B^2/8\pi << P + P_c$, and
the (Alfven) velocity of the magnetic scatterers is small relative
to the thermal gas
(e.g. Drury \& Falle 1986)
and these reasonable assumptions are adopted here.
We consider only non-relativistic bulk velocities $u \lta 0.1c$,
appropriate for our hotspot-driven flows. 
The hotspot acceleration $a_{hs}$ and cosmic ray luminosity 
density ${\dot S}_{hs} = d e_{c,hs}/dt$ 
are included as source terms in the equations above.

These equations are solved here using our own 
2D axisymmetric ZEUS-like hydrocode in cylindrical coordinates.
This code has been rather extensively checked 
in particular by exactly duplicating the shock 
structure of Jones \& Kang (1990) which is modulated
by diffusing cosmic rays.
In our codes the cosmic ray diffusion term
${\bf \nabla}\cdot (\kappa {\bf \nabla} e_c)$
is solved using operator-splitting and
fully implicit Crank-Nicolson differencing.
As in previous recent papers 
(Mathews \& Brighenti 2008; Mathews 2009),
we employ a cosmic ray diffusion coefficient 
that varies inversely with the local gas density, 
crudely assuming that the magnetic field also scales with density.
The diffusion coefficient depends on 
a single density parameter $n_{e0}$:
\begin{displaymath}
\kappa = \left\{
\begin{array}
{r@{\quad:\quad}l}
10^{30} ~{\rm cm}^2 {\rm s}^{-1} & n_e \le n_{e0} ~{\rm cm}^{-3} \\
10^{30}(n_{e0}/n_e) ~{\rm cm}^2 {\rm s}^{-1} & n_e > n_{e0} ~{\rm
  cm}^{-3}
\end{array} \right.
\end{displaymath}
However, as discussed earlier, it is very unlikely that 
the relativistic component in Cygnus A 
diffuses significantly during its short age $t_a = 10^7$ yr. 
For this reason we consider only a small density parameter
$n_{e0} = 6 \times 10^{-6}$ cm$^{-3}$ 
that effectively suppresses the role of CR diffusion in the solutions. 

In the calculations described here, 
the cosmic ray component is assumed not to lose energy 
by synchrotron or inverse Compton emission.
This is probably a reasonable assumption because,  
during the relatively short lifetime of cosmic rays in Cygnus A, 
the total cosmic ray energy is not expected to be greatly degraded 
by this emission. 
Radiative losses depend on the square of the particle energy, 
while, for the original particle energy spectrum in Cygnus A, 
$N(\gamma) \propto \gamma^{-p}$ with $p=2.2$ (Machalski et al. 2007),
most of the total cosmic ray energy $e_c$ is contained in 
low energy particles which emit very little radiation. 
However, the purely adiabatic evolution of the relativistic 
energy density $e_c$ does not mean that the total cosmic ray 
energy integrated over volume 
$E_c = \int e_c dV$ is conserved during the cocoon evolution.
Mechanical $PdV$ work done by (or on) the cosmic ray fluid 
as it interacts with the cluster gas can alter its total energy. 

\subsection{The Cluster Surrounding Cygnus A}

X-ray observations of the cluster gas surrounding 
Cygnus A are discussed in Smith et al. (2002) and subsequently 
modified by Wilson, Smith \& Young (2006).
The density and temperature profiles of this cluster are 
closely matched by those of Abell 478 which has been extensively 
observed by Vikhlinin et al. (2006).
Consequently we use the temperature fitting functions 
for Abell 478 provided by Vikhlinin et al. but reduce all temperatures 
by a factor 0.915 to exactly duplicate the innermost temperature 
measured near Cygnus A by Wilson, Smith \& Young, 
$kT = 4.60$ keV, at radius 32 kpc.
We assume that the cluster potential is described with an NFW 
potential for virial mass $1.25\times10^{15}$ $M_{\odot}$
and concentration $c = 7.61$ using Figure 8 of Vikhlinin et al. 
In addition, we assume that the central galaxy in the 
Cygnus A cluster is identical to M87 in the Virgo cluster, 
having an identical luminosity distribution 
and stellar mass to light ratio as found in M87 by 
Gebhardt \& Thomas (2009) and with a central black hole 
of mass $6.4 \times 10^9$ $M_{\odot}$ as determined by 
these authors.
Using the combined gravitational acceleration $g(r)$ 
of Abell 478 dark matter and stars plus black hole from M87, 
we integrated the equation 
for the cluster gas density in hydrostatic equilibrium,
\begin{equation}
{1 \over \rho}{ d \rho \over dr} =
- {1 \over T}\left[{dT \over dr} 
+ \left({\mu m_p \over k}\right)g \right]
\end{equation}
where $T(r)$ is the Abell 478 temperature profile adjusted as 
described above. 
Finally, beginning at some small radius, this equation was 
integrated using an initial density that exactly duplicates 
the gas density $\rho = 4.4 \times 10^{-26}$ gm cm$^{-3}$ 
($n_e = 2.27 \times 10^{-2}$ cm$^{-3}$) 
as observed by Wilson, Smith \& Young at radius 32 kpc 
from the center of Cygnus A. 
Cluster temperature and density profiles are illustrated
in Figure 2.

\subsection{Computation of the Phantom Hotspot}

We adopt a 2D cylindrical computational grid with 150 uniform zones 
$\Delta r = \Delta z = 0.5$ kpc extending out to 75 kpc in both 
directions, well beyond the Cygnus A cocoon at the present time.
Beyond 75 kpc, 50 additional zones 
having geometrically increasing sizes 
extend the grid to 0.8 Mpc in both directions. 
The phantom hotspot moves out along 
the symmetry $z$-axis at velocity $v_{hs}(t)$. 
In keeping with the kpc-sizes of observed hotspots, 
we consider phantom hotspots that extend 1kpc in the r-direction 
(two $j$ zones) and 0.5 kpc in the z-direction (one $i$ zone).
When the instantaneous phantom hotspot radius 
satisfies $z_{i-{1\over2}} < r_{hs}(t) < z_{i+{1\over2}}$, 
the innermost two  
$j$-zones at the $i$th grid along the $z$-axis 
are identified as the hotspot.
In updating the gas velocity during each 
computational time step $\Delta t$, 
the hotspot zones receive an additional acceleration 
in the $z$-direction 
\begin{equation}
a_{hs;i,j} = { \rho_{i,j} v_{hs}(t)^2 A_{i,j} 
\over \rho_{i,j} A_{i,j} \Delta z }
= {v_{hs}(t)^2 \over \Delta z}
\end{equation}
where $A_{i,j} = \pi(r_{j+1}^2 - r_{j}^2)$ 
is the area of the $j$th hotspot zone 
and $\Delta z = z_{i-1} - z_i$.
Time steps are chosen so that the phantom hotspot 
computation moves 
slowly along the z-axis, spending many time steps in 
each hotspot zone as it accelerates gas to $v_{hs}(t)$.
Consequently, each hotspot zone undergoes an additional 
operator-splitting step, 
\begin{equation}
{u_{i,j}}^{n+1} = 
\min[{u_{i,j}}^n + v_{hs}^2 \Delta t/ \Delta z , v_{hs}].
\end{equation}
During each time step $\Delta t = t^{n+1} - t^n$ hotspot zones also 
receive an additional increment of cosmic rays.
\begin{equation}
{e_{c;i,j}}^{n+1} = 
{e_{c,i,j}}^n + {\dot S}_{hs} \Delta t
\end{equation}
and ${\dot S}_hs \propto L_{cr}/V_{hs}$ (equations 4, 5 or 6)
where $V_{hs}$ is the total volume of all hotspot zones, 
assumed to be constant during the calculation.

\section{Computed Cocoons for Cygnus A}

We compare the distribution of gas and cosmic rays 
in three dynamical models, all at time $t_a = 10^7$ yrs 
when the hotspot has moved 60 kpc from the galaxy/cluster center. 
The total cosmic ray energy injected into the hotspot, 
$E_{0;cr} = t_a L_{cr} = 3.15 \times 10^{60}$ ergs, 
is the same in all models, but the mechanical work done by 
hotspot compression varies with the choice of $v_{hs}(t)$.

\subsection{Cocoon with  Uniform $v_{hs}$ and $\sigma_{hs}$}

Figure 3 shows the FR II cocoon at time $t_a = 10^7$ yrs 
for model 1 in which $\eta = \sigma_{hs} = 1$ (see eqn. 5) 
so that both the hotspot velocity 
$v_{hs} = r_a/t_a = 5870$ km s$^{-1}$ 
and cosmic ray production are constant along the 
hotspot trajectory (horizontal or $z$-axis). 
The upper panel shows in crossection 
white contours for the cosmic ray energy density $e_c(z,r)$ superimposed 
on a logarithmically-scaled image of the gas density $\rho(z,r)$. 
The very low density dark core of the cocoon is completely filled
with cosmic rays that define the radio lobe region. 
As in Figure 1, the cosmic ray contours in Figure 3 are 
closely spaced around the 
perimeter of the radio cavity where the relativistic and thermal 
gases meet in a contact discontinuity.
The hotspot is visible at the far right as 
a small completely white region 60 kpc along the $z$-axis. 
The cavity surface is disturbed near $(z,r) = (25,7.5)$ kpc, 
but very little gas flows across the radio lobe boundary.

The lower panel in Figure 3 shows 
an image of the bolometric thermal X-ray surface brightness 
$\Sigma_x = \int (\rho / m)^2 \Lambda d \ell$
integrated along the line of sight $\ell$ assumed to be 
perpendicular to the $z$-axis.
Here the bolometric emissivity 
$(\rho / m)^2 \Lambda(T,z_{\odot})$ erg cm$^{-3}$ s$^{-1}$ 
is evaluated assuming solar abundance. 
The contours in this panel outline the projected energy density 
$\int e_c d \ell$, giving a rough idea of the limits and appearance 
of the radio emission in projection if the cavity 
magnetic field were uniform. 

Figure 4 shows profiles of $\rho(z,0)$ (solid line) 
and $e_c(z,0)$ (dashed line) along the jet axis ($r=0$) 
at time $t_a = 10^7$ yrs.
The dramatic concentration of cosmic rays in the hotspot 
qualitatively resembles the Cygnus A X-ray and radio images (Fig. 1), 
but the computed hotspot is not bright in X-rays in Figure 3 
since we do not include synchrotron self-Compton or inverse 
Compton emission from the CMB. 
As expected, the gas density is compressed just ahead of the 
hotspot forming the apex of the cocoon bow shock.

The image in Figure 5 shows the temperature 
distribution $T(z,r)$ of the extremely hot, very low 
density thermal gas within the radio cavity. 
The thermal gas inside the radio lobe 
is evidently heated by multiple shocks produced 
by rapidly propagating waves in the relativistic gas 
trapped within the radio cavity walls. 
These low-amplitude waves are visible in Figure 4 and 
in the cavity contours in Figure 3. 
The temperature of essentially all of 
the heated thermal gas inside the X-ray cavity 
is relativistic, i.e. $T \gta 2m_ec^2/3k \approx 4\times10^9$ K, 
so its thermal properties and total energy are not accurately computed 
with the non-relativistic gas equations used here.
The acceleration of cluster gas to mildly relativistic velocities
inside the radio cavity is almost certainly a collisionless process 
because of the very low particle density. 

After $10^7$ yrs the total mass of thermal gas with $T > 10^9$ K, 
all of it inside the radio cavity, is $1.7\times10^8$ $M_{\odot}$. 
Most of this gas flows directly from the hotspot. 
For example, the mass of a cylindrical core 
through the undisturbed cluster gas 
having the same 1-kpc radius as the hotspot and length 
equal to $r_a = 60$ kpc is $1.7\times10^8$ $M_{\odot}$, 
essentially identical to the mass of ultra-hot cluster gas. 
The average density of gas with $T > 10^9$ K inside the 
radio lobe is $\langle n_e \rangle_{lobe} = 1.8\times10^{-4}$
cm$^{-3}$, which is slightly less than one percent of the density 
in the original cluster gas in the same region.
We suspect that the mass of ultra-hot cavity gas with $T > 10^9$ K 
depends on detailed grid-level assumptions about the 
exact spatial distribution of cosmic rays inside the hotspot. 
In our calculations we assume that cosmic rays are completely mixed 
with the thermal gas in the hotspot grid zones, 
but if they are not so efficiently mixed, the 
outflow of thermal gas from the hotspot 
and the mass of ultra-hot cavity gas might be lowered. 

In Figure 6 we show crossection profiles of the gas pressure $P$,
relativistic pressure $P_c$ and the gas density $n_e$ 
plotted perpendicular to the jet (or $z$) axis at 
$z = 10$ and 45 kpc, both at time $t_a = 10^7$ yrs. 
The total pressure $P + P_c$ is shown with a dotted line, 
indicating a cocoon overpressure several times larger than 
that in the original cluster gas at both locations 
along the jet axis. 
The total pressure in the radio lobe, 
$P + P_c \approx 10\times10^{-10}$ dynes cm$^{-2}$, 
is essentially constant along the jet axis 
(as in Fig. 4), consistent with an absence of 
significant mass backflow in the cavity region. 
In the lower panel of Figure 6 (at $z = 45$ kpc) notice that  
the dash-dot profile for the thermal gas density $n_e$ 
becomes very small within $r = 6$ kpc where the
gas pressure is still dominant and appreciable; 
this explains the dark, low-density transition 
region in Figure 3 that extends a little beyond the 
region of cosmic ray confinement, most visible at $z \gta 30$ kpc. 

The final drop in $P$ and $n_e$ in the Figure 6 panels 
occurs at the bow shock, broadened by artificial viscosity.
In blast waves of this type it is possible to 
determine the Mach number ${\cal M}$ 
from the Rankine-Hugoniot shock jump transition  
$P_2/P_1$ or $\rho_2/\rho_1$.
The gas pressure and density shock transitions 
in Figure 6 indicate 
${\cal M} \approx 1.8$ at both $z = 10$ and 45 kpc.
This Mach number ${\cal M} = v_{sh}/c_s \approx 2.0$ 
is confirmed by direct measurement of the velocity $v_{sh}$  
of the profiles during times 
near $t_a$ and the local cluster gas sound speed.
Wilson et al. (2006) estimate a somewhat smaller Mach number 1.3 
from X-ray observations of the cocoon shock in Cygnus A. 

While the cocoon in Figure 3 has many features in common
with Cygnus A, the overall shape of the computed cocoon shock 
and radio cavity are different. 
The cocoon shock in Figure 1 is uniformly convex everywhere, 
unlike the mildly concave shock surface in Figure 3 with 
a slope change near $(z,r) = (10.7,25)$ kpc.
In addition, the radio ``bridge'' region that connects
opposing hotspots in Cygnus A (e.g. Steenbrugge et al. 2010)
is more nearly cylindrical, unlike the quasi-conical 
radio cavity region shown in Figure 3.
Evidently, too many cosmic rays are produced 
in model 1 near the center at early times in the evolution.

Our computed cocoon receives some energy from hotspot acceleration 
and compression but the cosmic ray energy, 
$E_{cr} = L_{cr}t_a = 3.1 \times 10^{60}$ ergs, dominates. 
The allocations of cocoon energy 
into kinetic, thermal and cosmic rays at time $t_a = 10^7$  yrs 
are listed in Table 1. 
The change in potential energy at time $t_a$  
is negligibly small, but is expected to become larger at 
later times (Mathews \& Brighenti 2008).
It is seen that the relativistic cosmic ray energy has dropped 
to about a third of its original value because of $PdV$ work 
done on the surrounding cluster gas.
To explore the energy introduced by hotspot acceleration, 
we compute ``model 1nocr'' for the cocoon evolution 
with hotspot acceleration as in model 1 
but with no contribution from cosmic rays, $L_{cr}=0$. 
Figure 7, showing the cocoon density pattern 
$\rho(z,r)$ for model 1nocr at 
time $t_a$, is strikingly different from the top panel in 
Figure 3. 
The dark wake behind the moving hotspot along the $z$-axis 
is a region of only moderately higher temperature, 
$T \sim 15\times10^7$ K, 
about three times hotter than the original cluster gas.
The shock wave is outwardly convex everywhere in Figure 7 and 
encloses a 
much narrower cocoon along the (vertical) 
$r$-axis than in Figure 3.
The cocoon shock Mach numbers are also lower 
at corresponding places, 
${\cal M} = 1.17$ at $z = 10$ kpc and 
${\cal M} = 1.27$ at $z = 45$ kpc. 
The energetics for model 1nocr listed in Table 1 reveal that the 
cocoon kinetic and thermal energies are both very much 
less than in model 1. 
The kinetic energy generated by hotspot acceleration 
and compression 
in model 1nocr can be estimated by assuming that the total mass
of initial cluster gas contained 
in a cylinder of hotspot area out to 
the current hotspot radius $r_a = 60$ kpc, $M_{hscyl} = 
1.7 \times 10^8$ $M_{\odot}$,
is accelerated to velocity $v_{hs} = 5800$ km s$^{-1}$.
The resulting kinetic energy $0.5M_{hscyl}v_{hs}^2 = 0.057 \times
10^{60}$ ergs agrees very well with the 
relatively small total energy for model 1nocr in 
the final column of Table 1.
This confirms the critical importance of 
cosmic ray energy to the overall cocoon energetics in model 1 
and implies that the width and shape of the radio cavity 
is sensitive to the local rate that cosmic rays are processed 
in the moving hotspot.
The final column of Table 1 is the sum of the previous three
columns, $\sum_iE_i = E_{kin} + \Delta E_{th} + E_{cr}$,
which for model 1 
is very close to the total cosmic ray energy injected, 
$E_{cr,inj} = L_{cr}t_a = 3.1 \times 10^{60}$ ergs. 
This serves as a global check on the accuracy of the 
calculations and the energy check itself. 
Finally, we estimate the total energy inside 
the X-ray cavity in Figure 3;
the sum of the internal energy ${\bar P}V/(\gamma_c-1)$ and  
the work done to generate the cavity ${\bar P}V$ is 
$4{\bar P}V \approx 3.1 \times 10^{60}$ erg, 
which is very similar to 
the total cosmic ray energy injected $E_{cr,inj}$. 
Here we use the mean cluster 
gas pressure ${\bar P} = 1.1\times10^{-9}$
dy cm$^{-2}$ within $r_a$ and the volume of the computed cavity
$V = 2.4\times 10^4$ kpc$^3$.

\subsection{Cocoon Models with Varying Hotspot Speed or 
Cosmic Ray Luminosity}

The preceding discussion suggests that the shapes of the 
cocoon shock and radio cavity could be improved if 
the hotspot source produced fewer cosmic rays 
during its early evolution.
Such a reduction could be achieved 
if the cosmic ray luminosity of the hotspot 
were lower or if the 
the hotspot velocity $v_{hs}$ were larger at early times. 
In this section we adjust the hotspot parameters 
to explore these two limiting cases in more detail. 

First consider cocoon model 2 having a non-uniform hotspot luminosity 
$\sigma(\tau)$ with parameters
$\eta = 0.1$, $\tau_e = 1/3$, $\Delta \tau = 0.1$ 
and source term ${\dot S}_{hs}$ as in equation (6).
Recall that $\tau = t/t_a$ and that $\sigma(\tau)$ is normalized 
so that the total cosmic ray energy injected in the hotspot 
$E_{cr,inj} = L_{cr}t_a$ 
remains unchanged at time $t=t_a$ but is distributed 
differently during the hotspot evolution. 
The $\sigma(\tau)$ corresponding to these parameters 
is plotted in Figure 8. 
Figure 9 illustrates the density $\rho(z,r)$ image 
and cosmic ray energy density $e_c(z,r)$ contours for model 2 
and their projected counterparts at time $t_a$.
As expected, this solution has a radio cavity 
boundary and bow shock profile that are more in agreement 
with those of Cygnus A in Figure 1. 

However, the increasing hotspot luminosity we consider 
in model 2 is opposite to the decreasing luminosity observed in the 
combined sample of compact symmetric and FR II radio sources 
discussed by Jeyakumar \& Saikia (2000). 
Consequently, it is of interest to consider 
model 3 in which the hotspot cosmic ray luminosity 
remains constant during its short lifetime $t_a$, 
but its velocity $v_{hs}(\tau)$ decelerates with time. 
Using equation (1) for $v_{hs}(\tau)$, 
the decelerating hotspot reaches $r = r_a = 60$ kpc at time $t_a$ 
while producing the same total cosmic ray energy 
$E_{cr,inj} = L_{cr}t_a$.
For model 3 we explore the decelerating hotspot velocity 
$v_{hs}(\tau)$ illustrated in Figure 10 
which is characterized with 
parameters $v_0 = 20,000$ km s$^{-1}$ and $\tau_0=0.15$ 
and source term ${\dot S}_{hs}$ as in equation (5).
The resulting gas density $\rho(z,r)$ and 
relativistic energy density $e_c(z,r)$ plus projections 
are shown in Figure 11.
The less conical shape of the radio cavity 
and the overall convex shape of the cocoon shock in model 3
are closer to those observed in Cygnus A 
(top panel of Fig. 1). 
The evolution of FR II sources from more compact sources 
has been discussed by Kawakatu, Nagai \& Kino (2008)
who suggest hotspot deceleration in various evolutionary scenarios.

Finally we note that the mass of ultra-hot gas
($T > 10^9$ K) inside the radio cavities in models 2 and 3,
$1.5\times10^8$ and $1.4\times10^8$ $M_{\odot}$ respectively,
are comparable with that found in model 1 and are
nearly equal to the total gas mass in 
a column of original cluster gas containing all hotspot zones
within the current hotspot-center distance, $r_a = 60$ kpc.
The volumes of the radio cavities in models 2 and 3,
$2.2\times10^4$ and $2.8\times10^4$ kpc$^3$,
are also similar to that in model 1, 
$2.4\times10^4$ kpc$^3$.

\section{Discussion}

\subsection{Orientation of Cygnus A}

In this paper we assume with Smith et al. (2002)
that the image of Cygnus A in Figure 1 
lies entirely in the plane of the sky. 
However, the western radio jet appears more luminous than 
its eastern counterpart (lower panel of Figure 1); 
the radio flux density ratio of the W/E jets on kpc scales, 
$R = 2.6\pm1$ (Carilli et al. 1996), is uncertain but 
consistent with some Doppler boosting. 
It is likely therefore that the image in Figure 1 with 
major to minor axis ratio of the cocoon shock 
$R_{maj/max} = 2.2$ 
(Wilson, Smith, Young 2006) 
is somewhat foreshortened.
If so, the current distance of the hotspot from the 
galaxy center exceeds the apparent distance $r_a = 60$ kpc 
and the ratios of the major to minor axis for 
the cocoon shock that we calculate 
($R_{maj/max} = 2.1$, 3.0 and 2.4 in Figures 3, 9 and 11 
respectively) 
are systematically too small. 
In Figure 7 for model 1nocr the aspect ratio 
$R_{maj/max} = 3.6$ is very large, indicating that 
the non-uniform production of cosmic ray energy 
by the hotspot must be adjusted for any assumed 
foreshortening correction 
until $R_{maj/max}$, as well as the shape 
of the FR II X-ray cocoon shock and its radio lobe  
all come into agreement.

\subsection{Unexplained Features in the X-ray Image}

In addition to the purely thermal X-ray emission in our 
models, the X-ray image of Cygnus A shown in Figure 1 also 
includes synchrotron self-Compton (SSC) X-ray emission 
from the hotspots and from the X-ray cavity 
(radio lobe region). 
To access the visibility of the Cygnus A cavity from thermal 
X-ray emission alone, we show in Figure 12 
X-ray surface brightness scans perpendicular to the 
jet axis for our model 1.
Evidently the X-ray cavity visible in Figure 12 
is filled with SSC emission 
at a level that by coincidence approximately matches 
the maximum thermal X-ray surface brightness 
just outside the radio lobe boundary, 
concealing the expected cavity in thermal X-rays. 
For example, 
the sharp outer edges of the radio lobes in the lower panel of 
Figure 1 do not appear to correspond to changes in the 
X-ray surface brightness in the upper panel. (The shape 
of the leading edge of the radio lobe in Figure 1 is 
broader than in our calculations presumably because the 
jet direction in Cygnus A changes rather abruptly, depositing 
cosmic rays over a broader region.) 

We also note the bright, asymmetric thermal X-ray emission 
visible in Figure 1 distributed 
roughly perpendicular to the jet direction and extending 
$\sim20$ kpc from the center of Cygnus A 
mostly toward the east.
This prominent X-ray feature is 
unrelated to the cocoon evolution from 
cluster gas that we study here.
Instead, this irregular X-ray emission may 
be the expanded remnant of denser gas 
formerly located near the center of Cygnus A 
that was shocked and heated by AGN energy 
during the early stages of FR II jet development. 
There is considerable observational evidence 
at other wavelengths for cooler, high density 
gas extending several kpcs from the center of Cygnus A. 
Optical emission lines 
observed in Cygnus A 
are characteristic of warm gas at temperature $T\sim10^4$ K 
that may be photoionized and heated
by a hard UV to X-ray spectrum from the central AGN
(Osterbrock \& Miller 1975).
The total mass of warm gas is $\sim10^7$ $M_{\odot}$.
Line emission from the nuclear regions is observed 
to be significantly reddened by dust 
intrinsic to Cygnus A 
(Taylor, Tadhunter \& Robinson 2003). 
Near infrared observations by Wilman et al. (2000) 
detect rovibrational emission lines of H$_2$
as well as [FeII] and H recombination lines.
These lines are spatially extended by a few kpc  
and appear to come from different regions with 
complex velocity profiles 
having widths up to nearly $\sim500$ km s$^{-1}$
(Bellamy \& Tadhunter 2004).
Wilman et al. estimate the mass of molecular gas to 
be $10^8 - 10^{10}$ $M_{\odot}$. 
Soft X-ray absorption columns of ${\cal N} \sim 10^{23}$ cm$^{-2}$ 
are commonly observed in FR II sources with 
accompanying fluorescent Fe K$\alpha$ emission from 
cooler gas (Evans et al. 2006), 
so Cygnus A is not unusual in having massive central 
reservoirs of colder gas.
If the asymmetric display of thermal X-ray emission 
in Figure 1 results from colder gas that was 
shock-heated near the center 
of Cygnus A during the early development of its 
FR II event, as we suggest here, 
it is possible that the energy absorbed in heating and 
expanding this gas to its present position 
would reduce the width of the cocoon shock 
near the center computed in our model 1,
agreeing better with the observed cocoon shape and 
preserving the assumption of spatially uniform 
cosmic ray luminosity in the hotspot.

Finally we draw attention again to the faint but 
clearly visible X-ray ``jets'' in Figure 1, 
several kpc in width, 
extending along the major axis of the cocoon  
both east and west from the center of Cygnus A.
This surface brightness of this quasi-linear feature is 
irregular but has a cylindrical appearance overall.
It is difficult to understand how such a feature 
could be created by a radio jet. 
Fluorescent AGN X-ray emission from 
a radiating bi-cone along these opposing directions 
also fails unless the 
gas temperature in this region is very much less than 
a few keV.

\subsection{Components in the Radio Lobe Pressure}

The initial undisturbed cluster gas pressure varies 
across the Cygnus A cocoon from 
$20\times10^{-10}$ dynes cm$^{-2}$ 
at the cluster center to 
$1.9\times10^{-10}$ dynes cm$^{-2}$ at the hotspot 
radius $r_a = 60$ kpc. 
The initial pressure scale height in the cluster gas 
is approximately 
$r_P \propto P/\rho g \propto T/g$.
By comparison, inside the radio cavity 
the relativistic temperature $T_{lobe}$ and pressure scale 
height greatly exceed those in the cluster gas, 
$r_{Plobe} \propto T_{lobe}/g \gg r_P$, 
explaining why the (wave-averaged) pressure 
inside the radio cavity $P_{lobe}$ is nearly constant with 
cluster radius.
The total cavity pressure in models 1-3 discussed above is 
$P_{lobe} = 10.5$, 13.0 and 8.5 respectively 
in units of $10^{-10}$ dynes cm$^{-2}$, all approximately 
equal to the average pressure of the initial 
undisturbed cluster gas within the hotspot radius.

Recently Hardcastle \& Croston (2010) and Yaji et al. (2010) 
detected inverse Compton X-ray emission 
from Cygnus A. 
In many FR II sources the Compton emission  
results from electron interactions with cosmic background 
radiation (IC-CMB). 
However, since the synchrotron photon
density in Cygnus A exceeds that of the CMB, 
synchrotron self-Compton (SSC) X-ray emission dominates.
The additional information provided by 
Compton emission allows independent estimates  
of the energy density in the magnetic field 
$u_B = B^2/8\pi$ and in synchrotron-emitting electrons 
$u_e = m_ec^2\int \gamma n(\gamma) d\gamma$. 
The non-thermal synchrotron spectrum,
$F_{\nu}\propto \nu^{-\alpha}$, indicates that 
the electron energy distribution is also be a power law
$n(\gamma) \propto \gamma^{-p}$. 
The exponent $p \approx 2$
is expected from traditional shock-accelerated cosmic rays 
and is related to the spectral slope by 
$p = 2\alpha + 1$, suggesting 
$\alpha = 0.6 - 0.7$ for electrons when they are first 
injected into the radio cavity. 
Much of the non-thermal cavity energy 
is contained in low energy electrons 
and it often assumed that the integral for $u_e$ 
extends to some $\gamma_{min}$ far less than values 
$\gamma \sim 10^3 - 10^5$
that contribute to the observed radio spectrum.

If the total pressure in the radio lobe $P_{lobe}$ is known 
from X-ray observations or from dynamical models such as those 
presented here, 
it is possible to determine an  
energy density of non-thermal particles 
that matches the radio spectrum and, 
in combination with magnetic energy density (from IC), 
provides a pressure 
that must not exceed $P_{lobe}$. 
The parameters that define possible solutions for 
the particle energy densities are 
$p$, $\gamma_{min}$, and $\kappa$. 
the ratio of the energy density in non-radiating particles 
to radiating electrons. 
Both Hardcastle \& Croston (2010) and Yaji et al. (2010) 
find that the particle energy 
density in Cygnus A dominates over that of the magnetic 
field, $u_e \gg u_B$, so the interesting 
question is whether the observed synchrotron-radiating 
electrons provide 
enough pressure to support the cavity. 
The lobe pressure adopted by Yaji et al., 
$P_{lobe} \approx 20-40\times 10^{-10}$ dynes cm$^{-2}$ 
significantly exceeds those in our models while the 
value assumed by Hardcastle \& Croston, 
$P_{lobe} \approx 6\times 10^{-10}$ dynes cm$^{-2}$
is in better agreement.

With parameters 
$(p,\gamma_{min},\kappa) = (2,1,0)$ 
Hardcastle \& Croston find that the 
electron energy density necessary to match $P_{lobe}$ 
in Cygnus A produces inverse Compton emission that 
exceeds X-ray observations by factors of 2-5,
where the range reflects the rather large uncertainty 
in the X-ray Compton flux. 
Solutions (also with $u_e \gg u_B$) that match the 
Compton X-ray emission are possible if 
$(p,\gamma_{min},\kappa) = (2,1,1{\rm-}4)$ 
or 
$(p,\gamma_{min},\kappa) = (2.3,1,0)$.
Hardcastle \& Croston argue that 
$p=2$ is more physically acceptable for shock acceleration, 
and that a pressure component from additional 
non-radiating particles ($\kappa > 0$) contributes  
to the Cygnus A energy density, 
as in other FR II sources.
However, as discussed by Stawarz et al. (2007),
the physical conditions inside the Cygnus A hotspots
are not those of diffusive acceleration in a 
single traditional non-relativistic shock. 
Their infrared observations are consistent with a 
very flat low energy particle spectrum in the hotspot, 
$p \sim 1.5$,
and large fields, $\sim 200-300\mu$G, possibly indicating 
non-linear turbulent field amplification in 
the relativistic reverse shock.
Above $\nu \sim 1$GHz this flat hotspot spectrum steepens 
to $\alpha \gta 1$, 
suggesting to Stawarz et al. (2007) that two different 
accelerating mechanisms are involved and that only 
electrons emitting above $\sim1$GHz may be accelerated 
by a traditional diffusive-shock Fermi processes.

Notwithstanding all these complications,
the conclusion of Hardcastle \& Croston that 
some additional non-radiating pressure is required to support 
the Cygnus A cavity may well be correct. 
Adding relativistic shock-accelerated protons gains a factor 
of $\sim2$ but even this may not be sufficient. 
Another possibility is that there is an additional 
weakly relativistic electron population with energies 
$1 \lta \gamma \lta 10^3$ (or $10^9 \lta T \lta 10^{12}$ K) 
too low to contribute to currently available radio observations 
and therefore qualify as ``non-radiating''.
Note that these are similar to the temperatures of 
wave-shocked thermal gas inside our cavities, as in Figure 5.
Unlike diffusive shock acceleration where only about 10 percent 
of the thermal gas becomes relativistic, in our models 
the entire (small) mass of thermal gas inside the cavities 
is heated to relativistic temperatures presumably 
in a series of wave-driven shocks.
Since our computation of the dynamics of thermal gas 
is completely non-relativistic, we expect that the 
total pressure $P+P_c$ inside the cocoon cavity is 
correctly computed, but the 
pressure ratio $P_c/P$ inside the cavities is 
incorrect by a factor of order unity.
In a more realistic calculation that accurately describes 
the shock-heating transition of thermal gas to 
relativistic temperatures, we expect that
that some memory of the non-relativistic Maxwellian peak 
would be retained with a spectrum 
$n_{Max}(\gamma)$ 
quite unlike the power law spectrum resulting from traditional
diffusive shock acceleration.
The collisionless nature of cavity shocks introduces 
additional complications, but it is possible that 
shock-heated thermal gas provides an additional radio lobe pressure 
that is not apparent in the observed radio spectrum, 
even at the lowest currently feasible radio frequencies.
As discussed earlier,
the presence of a small mass of ultra-hot thermal gas in our 
computed cavities may be a model-dependent result 
from our assumptions that 
the FR II jets carry very few non-relativistic particles and  
cosmic rays and thermal gas fully mix  
before leaving the hotspot.
In any case, it is essential that wave activity or  
coherent fluid motions within the cavity 
do not disrupt the gradient of radio-synchrotron 
spectral steepening and electrons ages 
observed along the jet axis 
in the radio lobes of many FR II sources including Cygnus A;
some collisionless wave-damping mechanism may be required 
to accomplish this. 

\section{Conclusions}

We describe computations of 
the gas-dynamical evolution of FR II radio cocoons in galaxy clusters 
and, evidently for the first time, also include the self-consistent 
dynamical evolution of the relativistic synchrotron-emitting plasma.
Cluster and FR II parameters are chosen to approximately 
match those of Cygnus A, the most intensively observed and 
studied FR II source.
Our calculation is based on the hypothesis that the dynamics 
and energy content of FR II cocoons and non-thermal radio sources 
have their origin in the hotspots, 
where FR II jets first encounter the cluster gas.
Our hotspot-oriented computations allow us 
to avoid detailed calculations of the dynamical structure 
and particle content of the FR II jets 
about which very little is known
(e.g. Kataoka et al. 2008).
The strong reverse shock that forms at the inner 
hotspot boundary as a result of the jet impact 
imparts a momentum that drives the bow shock surrounding 
a cocoon of shocked cluster gas. 
To simulate this momentum transfer, we consider a phantom 
hotspot that moves out in some prescribed fashion into 
the computational grid along the jet axis, 
marking the grid zone which is 
compressed and accelerated in the reverse shock. 
However, the total thermal and kinetic 
power delivered by the hotspot is only a few percent of the 
power $\sim10^{46}$ erg s$^{-1}$ flowing from the 
hotspot in relativistic cosmic ray particles including the electrons 
that emit the observed radio radiation.
The cosmic ray energy introduced inside the hotspot is a combination
of relativistic particles arriving in the jet at the reverse 
shock and additional cosmic rays created in this shock 
by the diffusive Fermi mechanism.

In view of the domination of relativistic energy 
in forming the Cygnus A cocoon,  
we do not explicitly consider a 
non-relativistic mass flux from the jet into the hotspot.
As a result, in our calculations 
no mass accumulates in the high 
pressure region near the hotspot 
which in many previous FR II calculations drives 
a turbulent backflow near the jet in the opposite direction. 
Such backflows would disrupt 
radial gradients of radio-synchrotron ages and spectral steepening 
observed in FR II radio lobes and could generate 
irregularities in the radio cavity boundaries that 
are also not observed.

For computational simplicity we represent 
the cosmic ray energy only with its relativistic 
energy density $e_c$ ergs cm$^{-3}$  
and the corresponding pressure $P_c = (\gamma_c-1)e_c$ 
with $\gamma_c = 4/3$. 
We do not explicitly consider the energy spectrum of 
cosmic ray particles nor the physical nature of these particles 
which could be any combination of electrons and protons. 
In keeping with the small observed magnetic fields in Cygnus A
and other similar FR II sources, 
we assume that the magnetic energy density is everywhere
considerably less than the sum of the local thermal and
relativistic energy densities. 
However, a modest magnetic field that is frozen into the 
thermal cluster gas is essential for thermal and 
relativistic fluids to exchange momentum and respond 
to the total pressure gradient. 
With these simplifying assumptions, 
it is not possible to compute the synchrotron emission 
and radio spectrum since the energy spectrum of the electrons 
and the field strength are unspecified, 
but the radio lobe region is clearly defined by 
cavities in the hot gas displaced by the
relativistic fluid component. 
Finally, we assume that the relativistic and thermal fluids
interact and share energy adiabatically with no 
radiative losses due to thermal X-ray or 
synchrotron radio emission. 
This assumption is reasonable because of the short age 
of Cygnus A.

We regard the approximate synchrotron decay age 
observed in the radio lobes of Cygnus A, 
$t_a = 10^7$ yrs, as the dynamical age of the entire cocoon. 
The current projected distance of the hotspots from the center of 
Cygnus A, $r_a = 60$ kpc, is assumed to be identical with the 
actual physical distance, i.e. we imagine that the 
Cygnus A cocoon is essentially in the plane of the sky. 

An initial calculation in which the hotspot velocity 
and its cosmic ray luminosity $L_{cr} = 10^{46}$ erg s$^{-1}$ 
are assumed to be constant with time 
reproduces all the essential features of Cygnus A 
and those of FR II sources in general. 
The radio lobe filled with cosmic rays is extended along 
the jet direction and (for Cygnus A) 
merges at the origin with an identical 
mirror imaged counterlobe, forming a single extended 
``bridge'' of radio emitting cosmic rays 
along the jet-counterjet axis. 
The bow shock produced by the kpc-sized phantom hotspot 
encloses a cocoon of shocked gas that is comparable 
in size and aspect ratio to that in the X-ray image of Cygnus A.
The extended radio lobe confines the relativistic particles 
within the cocoon 
and at the same time displaces the shocked cluster gas in 
a sharp contact discontinuity similar to that observed 
in Cygnus A at radio frequencies. 
The energy budget is dominated by the internal 
energy of the relativistic component.
When an otherwise identical calculation 
is performed without cosmic ray injection in the hotspot,
the total kinetic and thermal energy received by the cluster gas 
is only a few percent of $t_aL_{cr}$.

The high pressure of cosmic rays received and generated 
by the jet at the hotspot causes both cosmic rays
and thermal gas to flow roughly transverse to the jet,
helping to widen the cocoon and strengthen the bow shock that 
encloses the cocoon.
As observed, the relativistic cosmic rays are 
confined inside elongated radio lobes.
However, we assume that 
the cosmic rays and thermal gas inside the hotspots 
are mixed and flow together into 
the apex of the radio cavity which contains a small 
mass of thermal gas.
The small mass of thermal gas that enters the radio lobe 
from the hotspot is shocked by high velocity waves to pressures that 
are comparable with that in the relativistic gas. 
This results in extremely high relativistic temperatures 
in the thermal gas that are not accurately calculated 
with our current non-relativistic hydrocode.
Nevertheless, the radially uniform pressure inside the 
Cygnus A radio lobe, 
about 2-3 times larger than the pressure at mid-lobe 
in the initial undisturbed cluster gas, should be accurately computed.
The total energy in our computations is conserved to a few percent.

We describe two additional computations in which fewer cosmic rays 
are introduced at early times in the Cygnus A evolution,
while keeping the total cosmic ray energy ejected by the hotspot 
and its average velocity unchanged. 
In these computations the shapes of the radio lobe and cocoon shock 
after $t_a = 10^7$ yrs are significantly improved. 
We considered two ways of achieving this early cosmic ray reduction: 
first by simply adjusting downward the cosmic ray hotspot luminosity 
at early times and 
second by assuming that the hotspot luminosity is constant 
but the hotspot velocity decelerates with time. 
The improved results of these two calculations are similar.
However, a uniform production of cosmic rays 
in the moving hotspot may still be possible if a significant 
fraction of the energy released by the hotspot during its
early evolution is absorbed in heating and accelerating 
cold gas away from the center of Cygnus A, 
producing the asymmetric emission observed in thermal X-rays.

\vskip.1in
\acknowledgements
Studies of the evolution of hot gas in elliptical galaxies
at UC Santa Cruz are supported by NSF and NASA grants 
for which we are very grateful.

%

\clearpage

\begin{deluxetable}{lcccccc}
\tabletypesize{\scriptsize}
\tablecolumns{6}
\tablewidth{11cm}
\tablecaption{COCOON HOTSPOT EVOLUTION AND ENERGIES
AT TIME $t_a = 10^7$ YRS}
\tablehead{
\colhead{model} &
\colhead{$v_{hs}$\tablenotemark{a}} &
\colhead{$\sigma$\tablenotemark{b}} &
\colhead{$E_{kin}$\tablenotemark{c}} &
\colhead{$\Delta E_{th}$\tablenotemark{c}} &
\colhead{$E_{cr}$\tablenotemark{c}} & 
\colhead{$\sum E_i$\tablenotemark{c}} \cr
\colhead{} &
\colhead{(km s$^{-1}$)} &
\colhead{} &
\colhead{($10^{60}$ erg)} &
\colhead{($10^{60}$ erg)} &
\colhead{($10^{60}$ erg)} &
\colhead{($10^{60}$ erg)} \cr
}
\startdata
1 & 5866. & 1.0 & 0.608 & 1.482 & 0.893 & 2.98 \cr
1nocr\tablenotemark{d} & 5866. & 1.0 & 0.015 & 0.042 & 0 & 0.06 \cr
2 & 5866. & $\sigma(t)$\tablenotemark{e} & 0.609 & 1.497 & 1.022 &
3.13 \cr
3 & $v_{hs}(t)$\tablenotemark{f} & 1.0 & 0.543 & 1.308 & 1.269 & 
3.15 \cr
\tablenotetext{a}{Hotspot velocity.}
\tablenotetext{b}{Time variation of hotspot cosmic ray source.}
\tablenotetext{c}{Cocoon energies at time $t_a$: $E_{kin}$, kinetic;
$\Delta E_{th}$, thermal
(with original cluster energy subtracted);
$E_{cr}$, cosmic ray. Final column is the sum of all three energies. 
All energies refer to one hemisphere.}
\tablenotetext{d}{Model 1 with no hotspot cosmic rays.}
\tablenotetext{e}{Variable hotspot cosmic ray source:
$\eta = 0.1$, $\tau_e = 0.3333$, $\Delta \tau = 0.15$.}
\tablenotetext{f}{Variable hotspot velocity:
$v_0 = 2\times10^4$ km  s$^{-1}$, $t_0 = 1.5\times10^6$ yrs.}
\enddata
\end{deluxetable}

\clearpage

\begin{figure}
\centering
\hspace{1.08cm}\includegraphics[width=4.4in]
{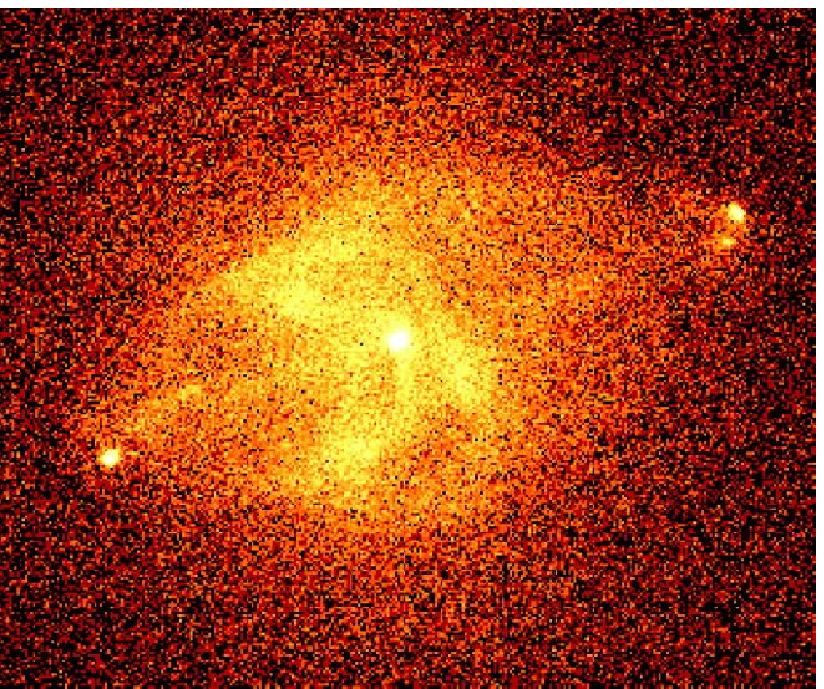}
\includegraphics[width=5.in]
{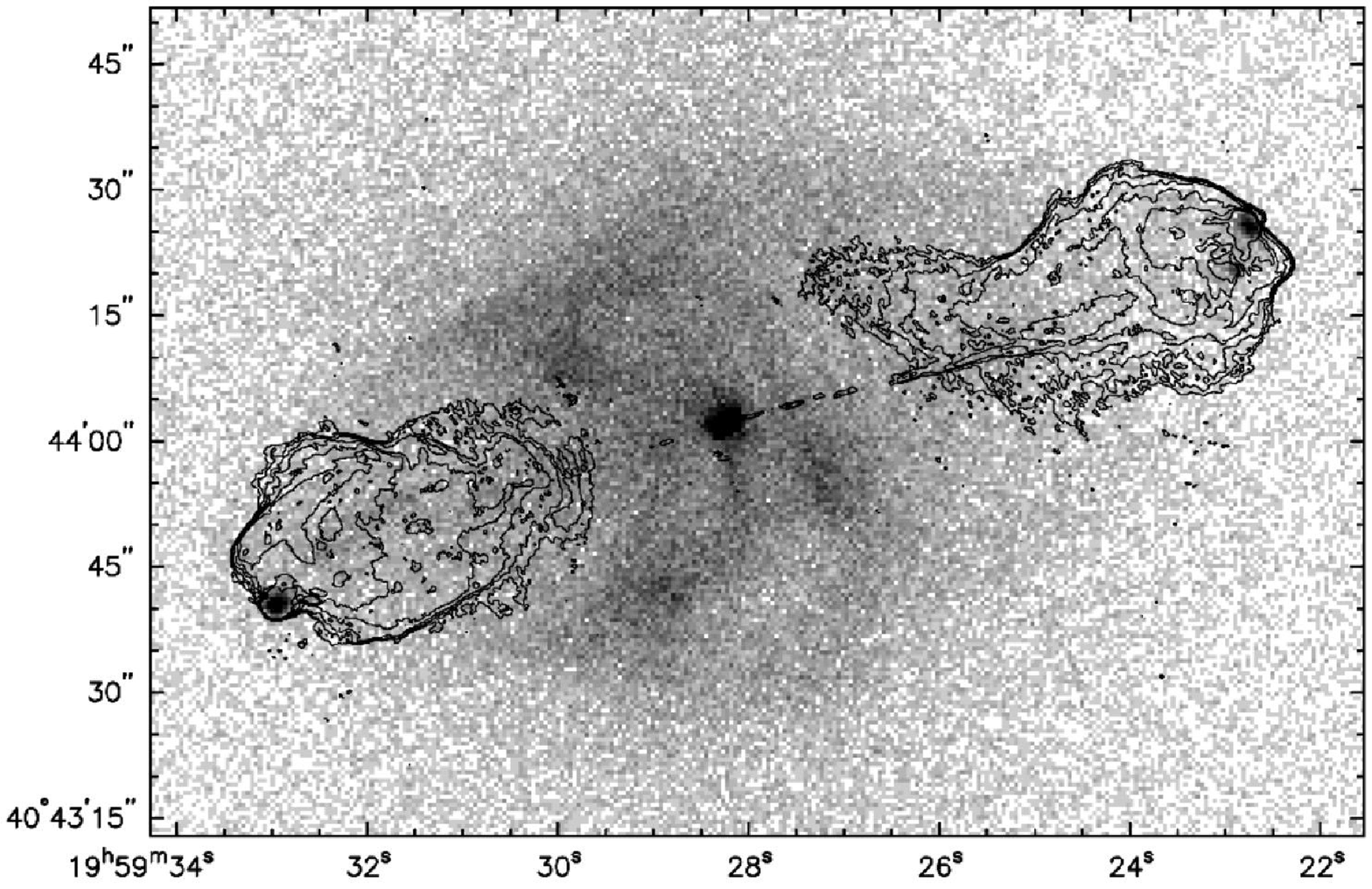}
\vskip0.3in
\caption{
{\it Top:} Chandra image of Cygnus A is 150 kpc
wide (1'' = 1 kpc).
Two oppositely-directed jets create a football-shaped
shock wave enclosing a cocoon of shocked gas.
{\it Bottom:} Same image with VLA radio contours at 5GHz.
(Wilson et al. 2006)
}
\label{f1}
\end{figure}

\clearpage

\begin{figure}
\centering
\includegraphics[width=5.in,angle=270]{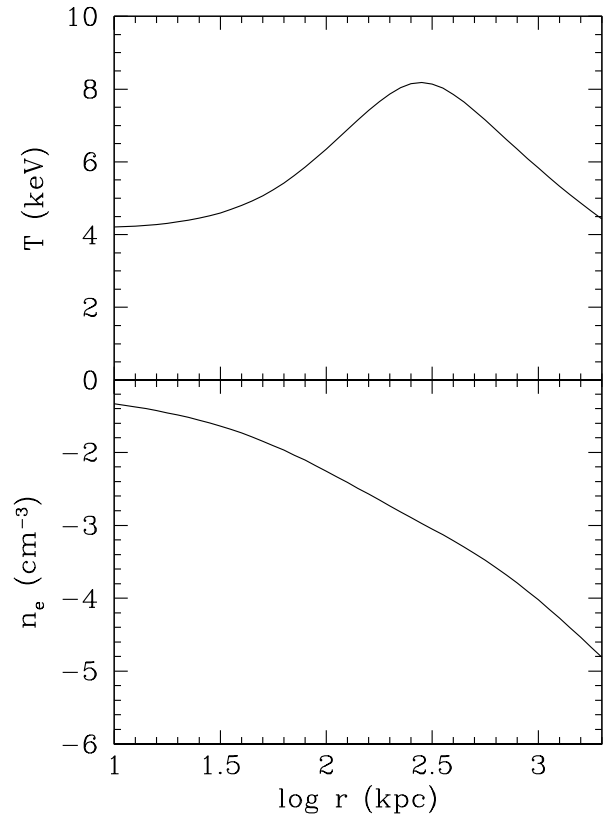}
\caption{
Adopted gas temperature and density profiles for the cluster 
surrounding Cygnus A.
}
\label{f2}
\end{figure}

\clearpage

\begin{figure}
\vskip-.3in
\centering
\includegraphics[width=7.in]{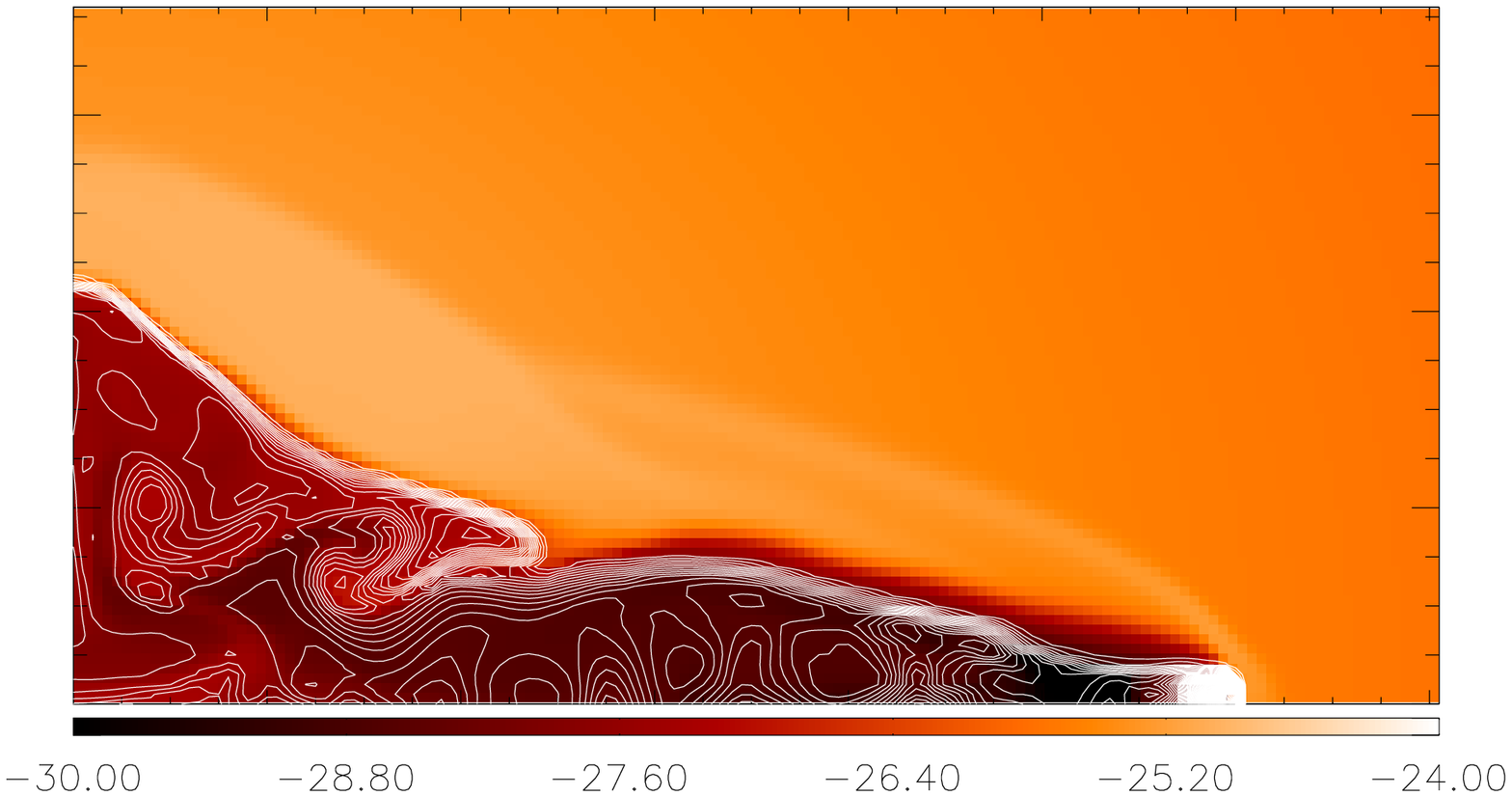}
\vskip0.15in
\includegraphics[width=7.in]{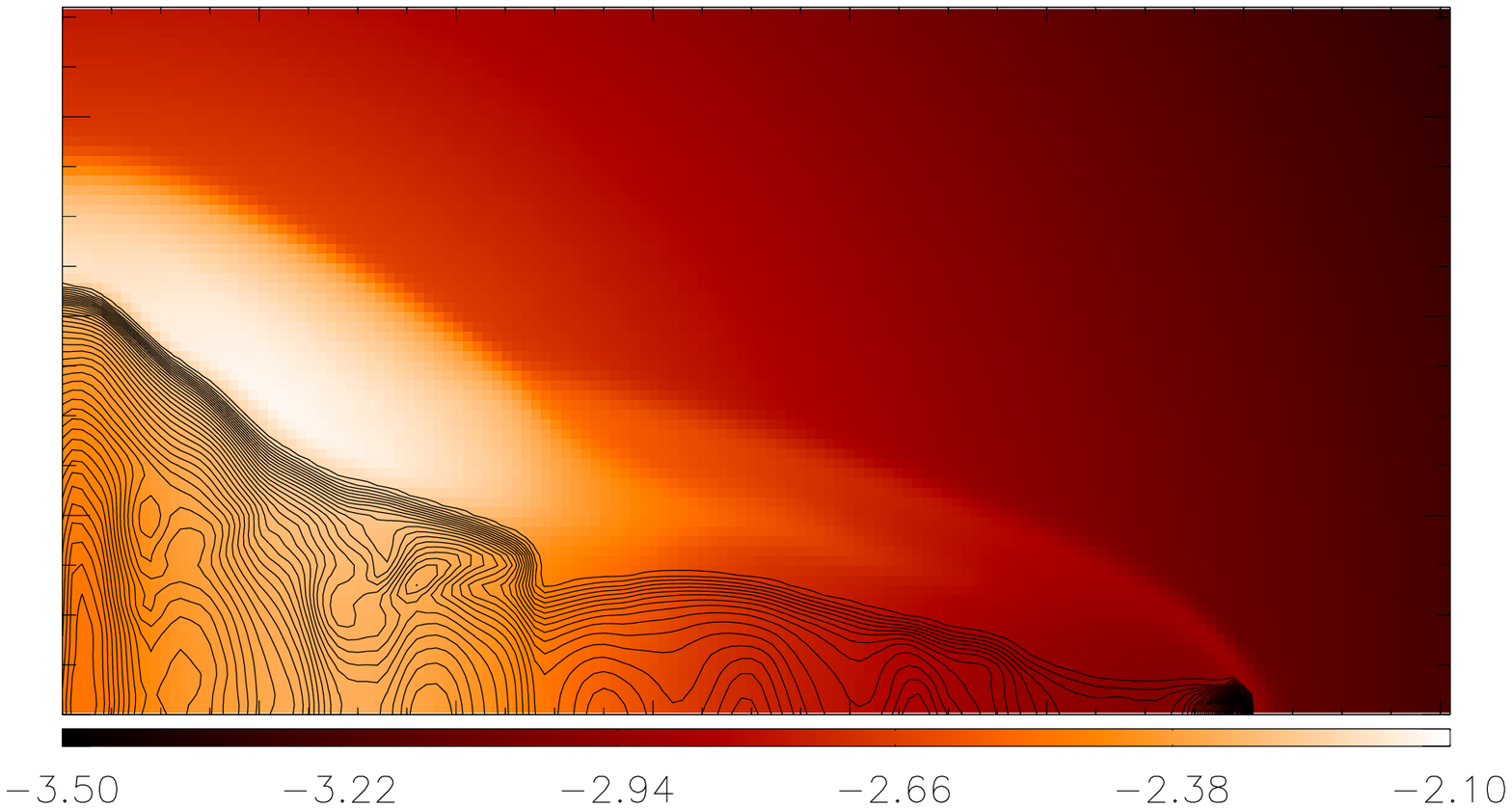}
\vskip.15in
\caption{
Model 1 at time $t_a = 10^7$ yrs. 
{\it Top}: Contours of $e_c(z,r)$ superimposed on image of 
$\log\rho(z,r)$ shown with values indicated in the colorbar.
{\it Bottom}: Image of the projected bolometric X-ray 
and emission contours of the  
projected cosmic ray energy density $\int e_c d\ell$. 
The horizontal $z$-axis and the vertical $r$-axis are marked in
(unlabeled) ticks ever 2 kpc with larger ticks every 10 kpc; 
both panels are $70 \times 35$ kpc.
The relativistic cosmic rays are confined in an elongated 
cavity in the shocked hot gas which is enclosed within 
the cocoon bow shock.
}
\label{f3}
\end{figure}

\clearpage

\begin{figure}
\centering
\includegraphics[width=5.in]{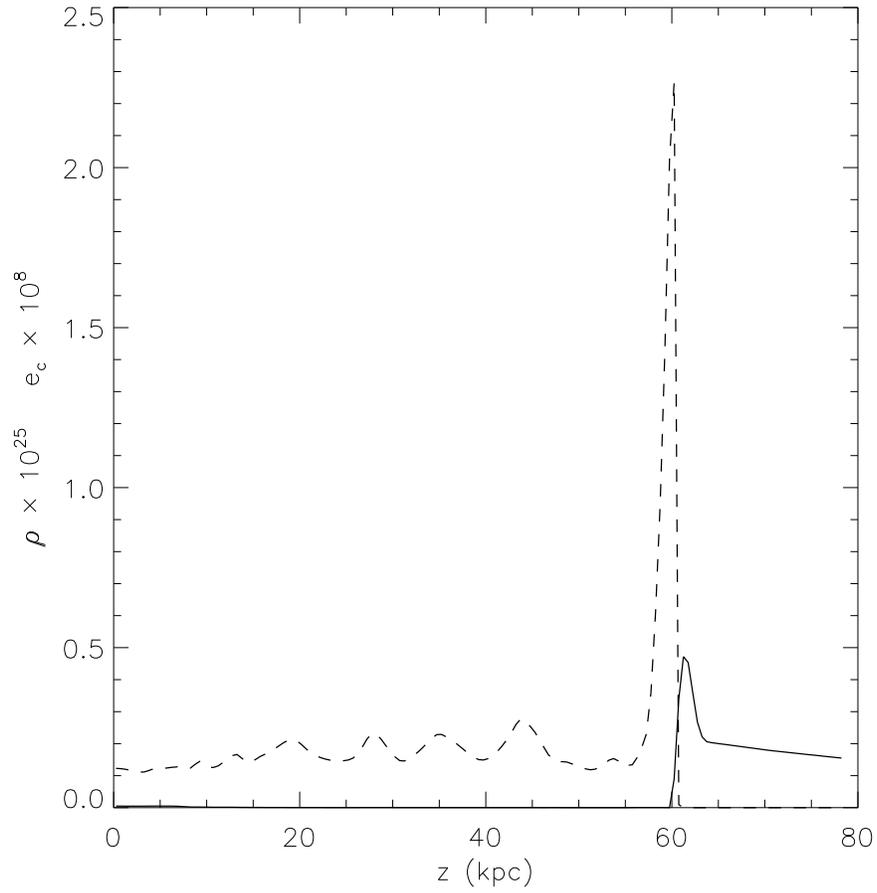}
\vskip0.5in
\caption{
Profiles of $10^{25}\rho(z,0)$ gm cm$^{-3}$ (solid line)
and $10^8e_c(z,0)$ erg cm$^{-3}$ (dashed line) 
along the jet axis ($r=0$) for Model 1 
at time $t_a = 10^7$ yrs.
}
\label{f4}
\end{figure}

\clearpage

\begin{figure}
\centering
\includegraphics[width=7.in]{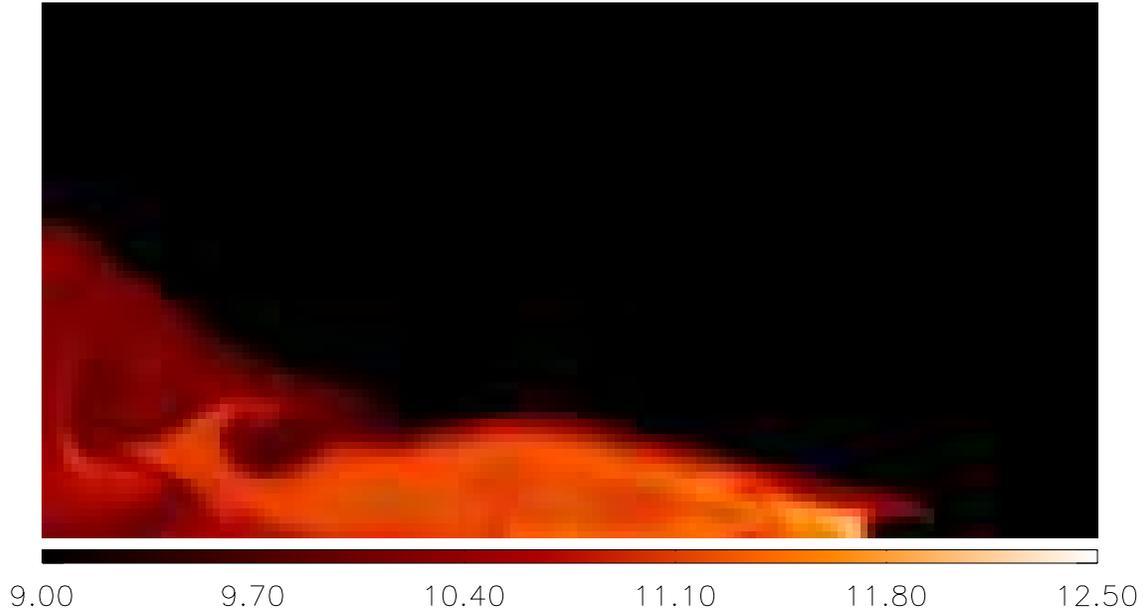}
\vskip0.5in
\caption{
Image of $\log T$ (K) of the very hot 
(thermal) gas within the X-ray cavity 
on the same $70 \times 35$ kpc spatial scale as Figure 3. 
The hottest gas $\log T \sim 11.5-12$ is in the wake just behind 
the hotspot and the temperature decreases to 
$\log T \sim 9.5-10.5$ closer to the cluster center where it 
is being mixed with thermal gas flowing into the cavity near 
$(z,r) = (25,7.5)$ kpc.
}
\label{f5}
\end{figure}

\clearpage

\begin{figure}
\centering
\includegraphics[width=5.in]{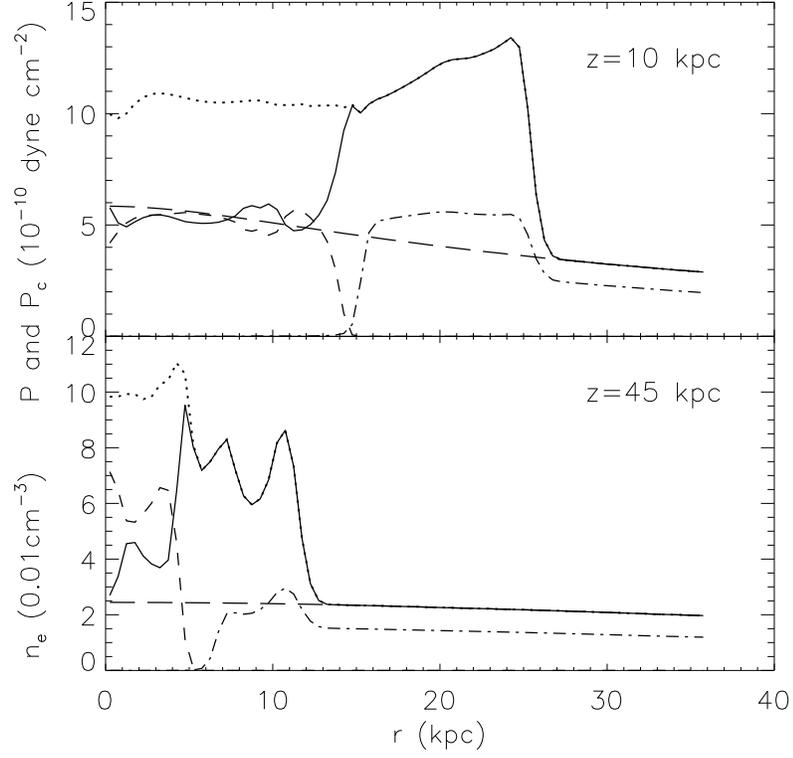}
\vskip0.5in
\caption{
Crossection profiles at time $t_a = 10^7$ yrs 
of the pressures $P$ (solid lines) and
$P_c$ (dashed lines), both in units of $10^{-10}$ dynes cm$^{-2}$, 
and the gas density $n_e$ (dash-dot lines) in units of 0.01cm$^{-3}$. 
The total pressure $P + P_c$ is shown with a dotted line.
The pressure of the initial undisturbed cluster gas inside 
the cocoon is shown with long dashed lines.
Profiles are perpendicular to the $z$-axis at
$z = 10$ and 45 kpc. 
}
\label{f6}
\end{figure}

\clearpage

\begin{figure}
\centering
\includegraphics[width=7.in]{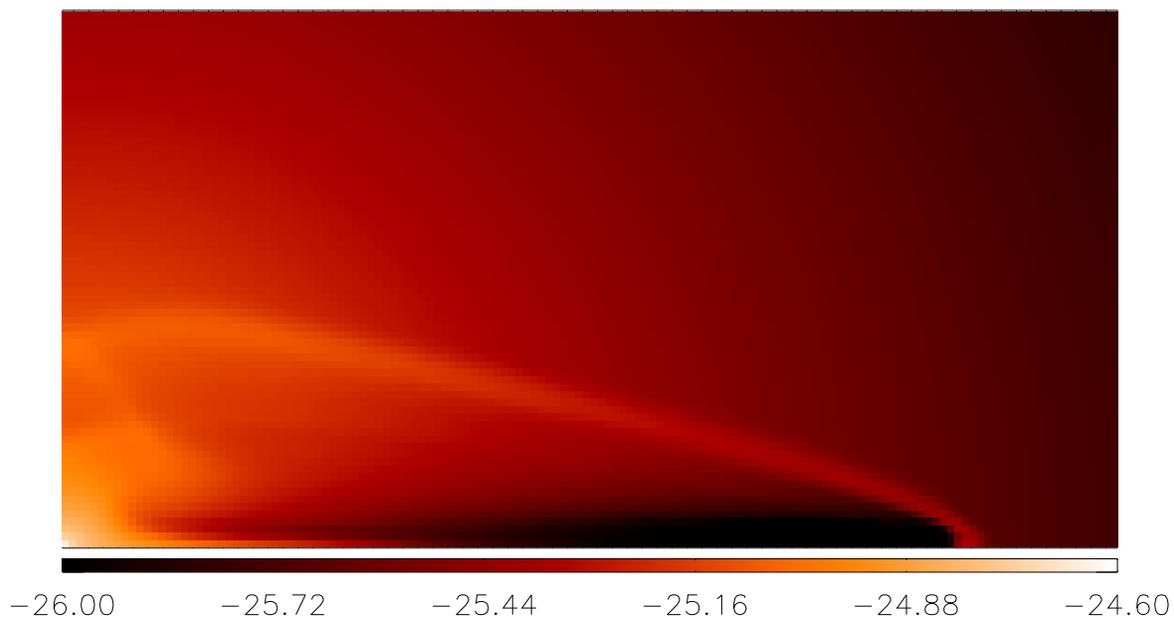}
\vskip0.5in
\caption{
Image of $\log \rho(z,r)$ at time $t_a = 10^7$ yrs for model 1nocr which 
is identical to model 1 but with no cosmic rays produced in the 
hotspot. The image is $70 \times 35$ kpc
}
\label{f7}
\end{figure}

\clearpage

\begin{figure}
\centering
\includegraphics[width=5.in]{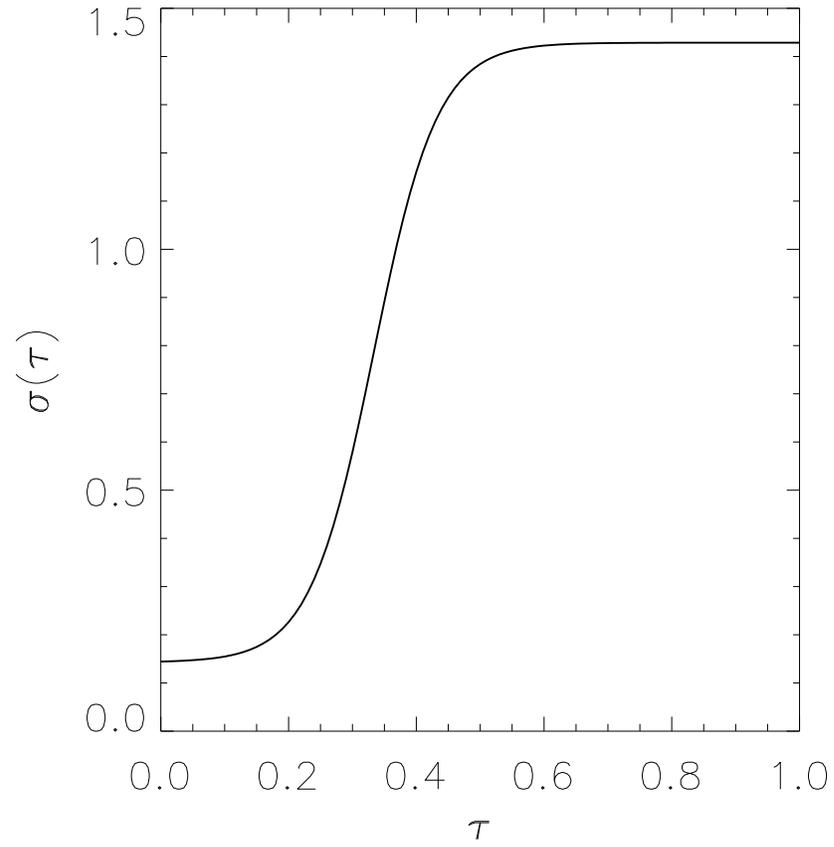}
\vskip0.5in
\caption{
Variation of the intrinsic hotspot cosmic ray luminosity 
factor $\sigma(\tau)$ with parameters
$\eta = 0.1$, $\tau_e = 0.3333$ and $\Delta \tau = 0.1$.
}
\label{f8}
\end{figure}

\clearpage

\begin{figure}
\centering
\includegraphics[width=7.in]{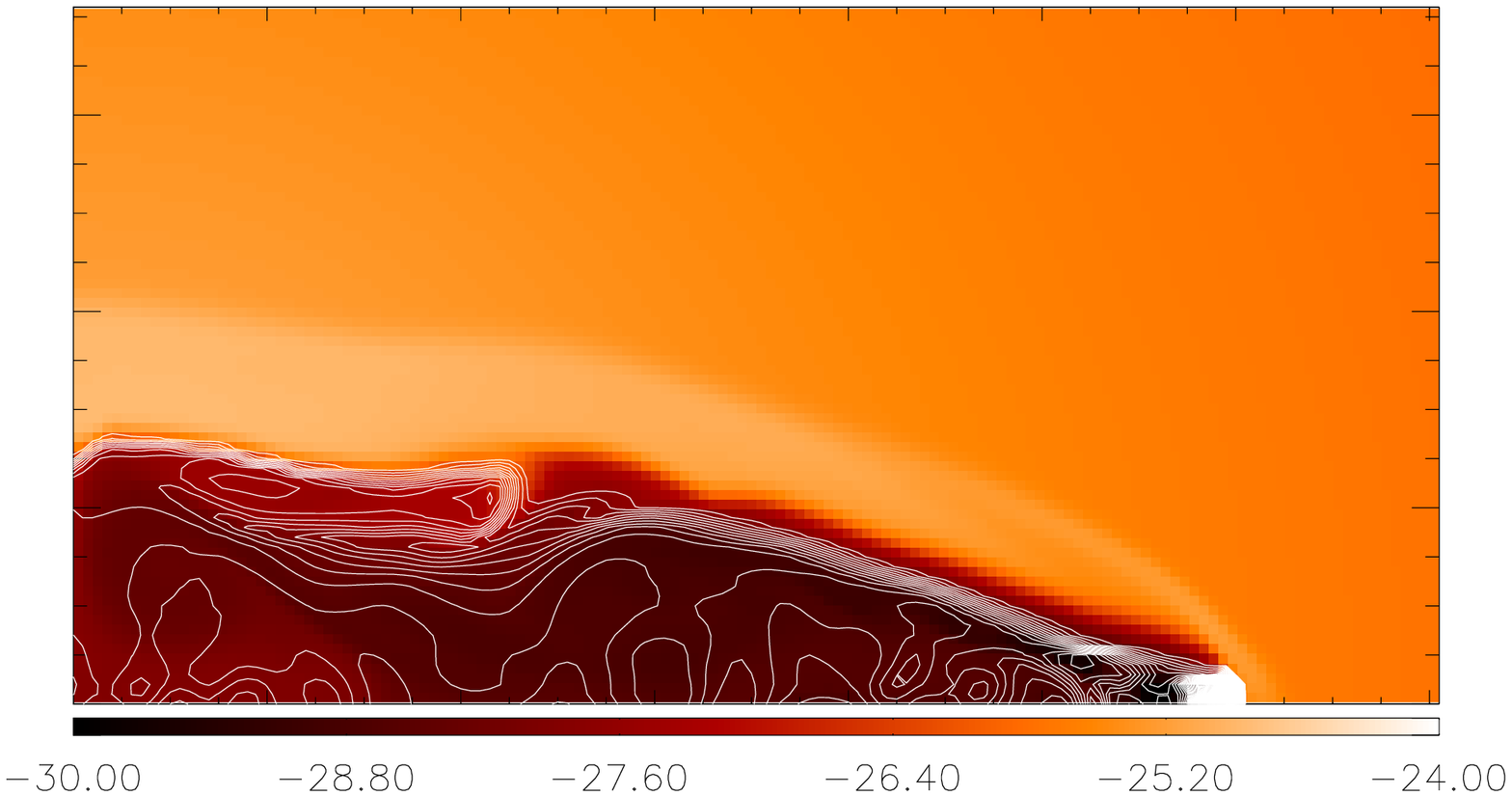}
\vskip.2in
\includegraphics[width=7.in]{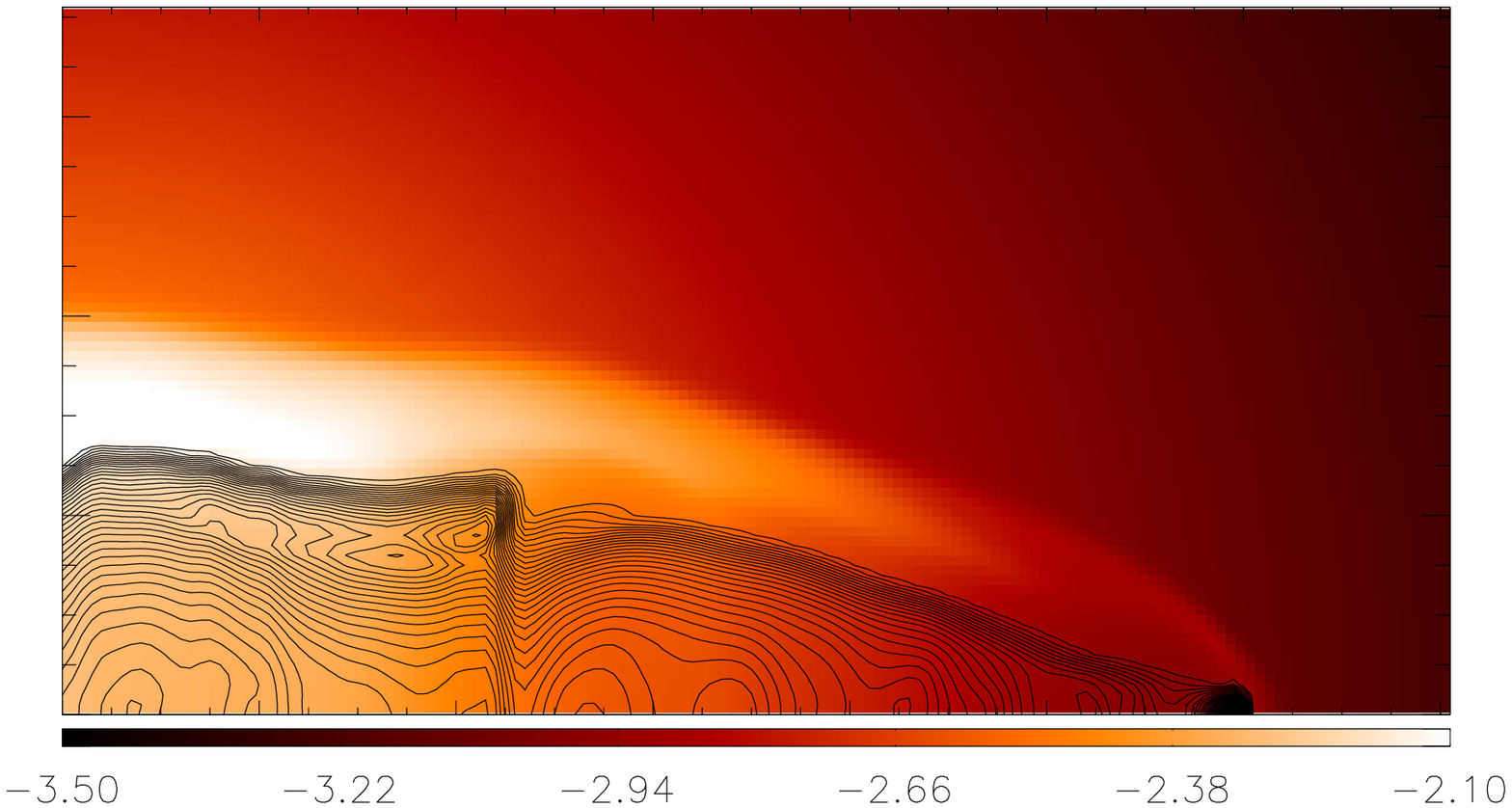}
\vskip0.5in
\caption{
Model 2 at time $t_a = 10^7$ yrs. 
{\it Top}: Contours of $e_c(z,r)$ superimposed on image of
$\log\rho(z,r)$ shown with values indicated in the colorbar.
{\it Bottom}: Image of the projected bolometric X-ray emission
and contours of the 
projected cosmic ray energy density $\int e_c d\ell$.
Both panels have dimensions $70 \times 35$ kpc.
}
\label{f6}
\end{figure}

\clearpage

\begin{figure}
\vskip3.in
\centering
\includegraphics[bb=250 216 422 769,scale=.8,angle=270]{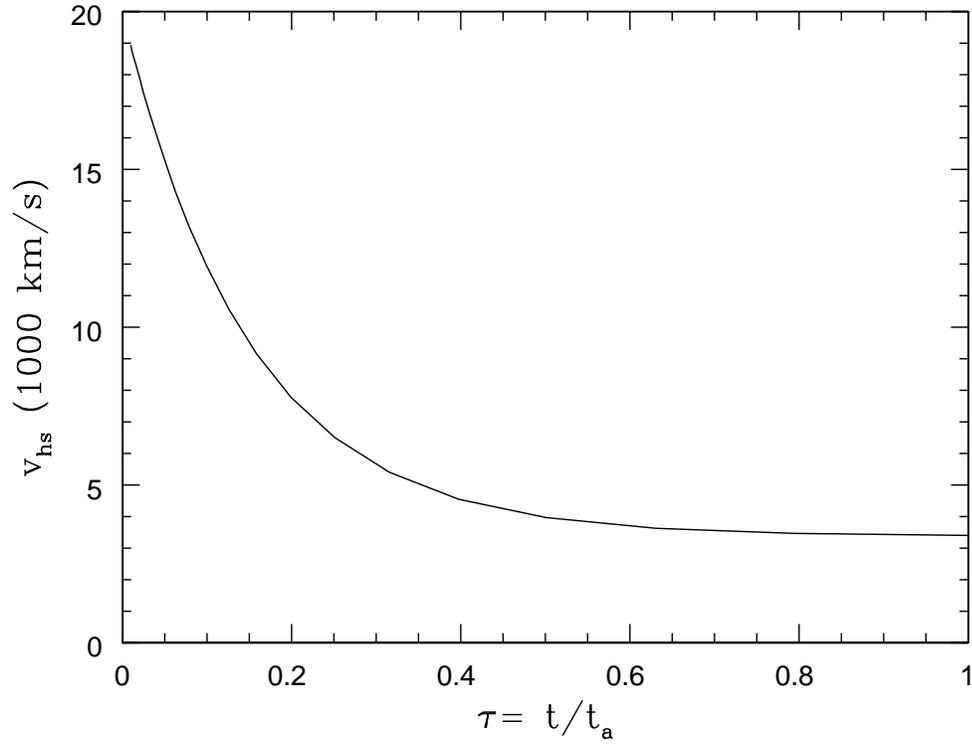}
\vskip0.5in
\caption{
Variation of the hotspot velocity $v_{hs}(\tau)$ with 
dimensionless time $\tau=t/t_a$ with parameters
$v_0 = 2\times10^4$ km  s$^{-1}$ and $t_0 = 1.5\times10^6$ yrs.
}
\label{f10}
\end{figure}

\clearpage

\begin{figure}
\centering
\includegraphics[width=7.in]{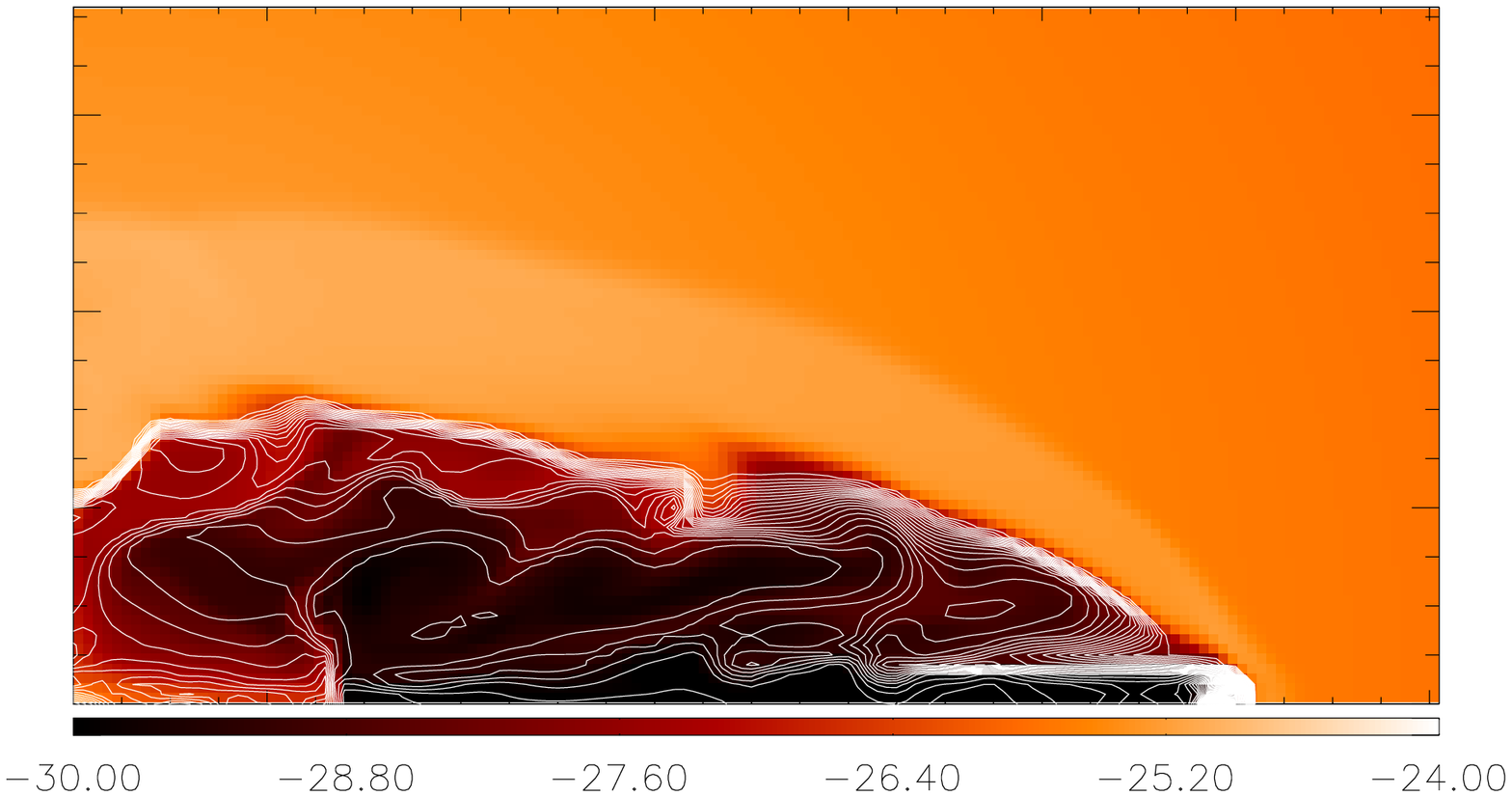}
\vskip.2in
\includegraphics[width=7.in]{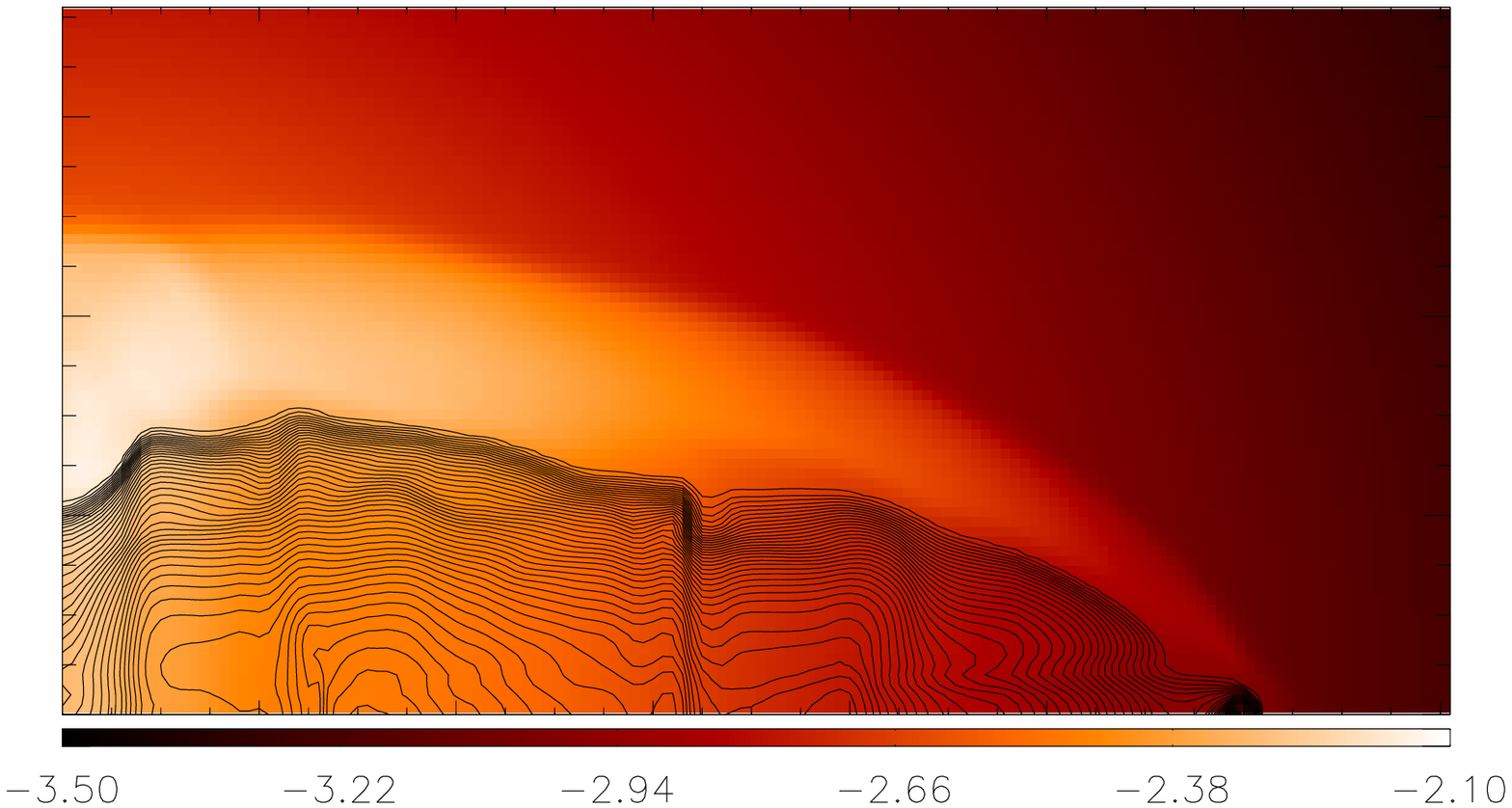}
\vskip0.5in
\caption{
Model 3 at time $t_a = 10^7$ yrs. 
{\it Top}: Contours of $e_c(z,r)$ superimposed on image of
$\log\rho(z,r)$ shown with values indicated in the colorbar.
{\it Bottom}: Image of the projected bolometric X-ray emission
and contours of the
projected cosmic ray energy density $\int e_c d\ell$.
Both panels have dimensions $70 \times 35$ kpc.
}
\label{f11}
\end{figure}

\clearpage

\begin{figure}
\vskip0.5in
\centering
\includegraphics[width=6.in,scale=.8,angle=0]{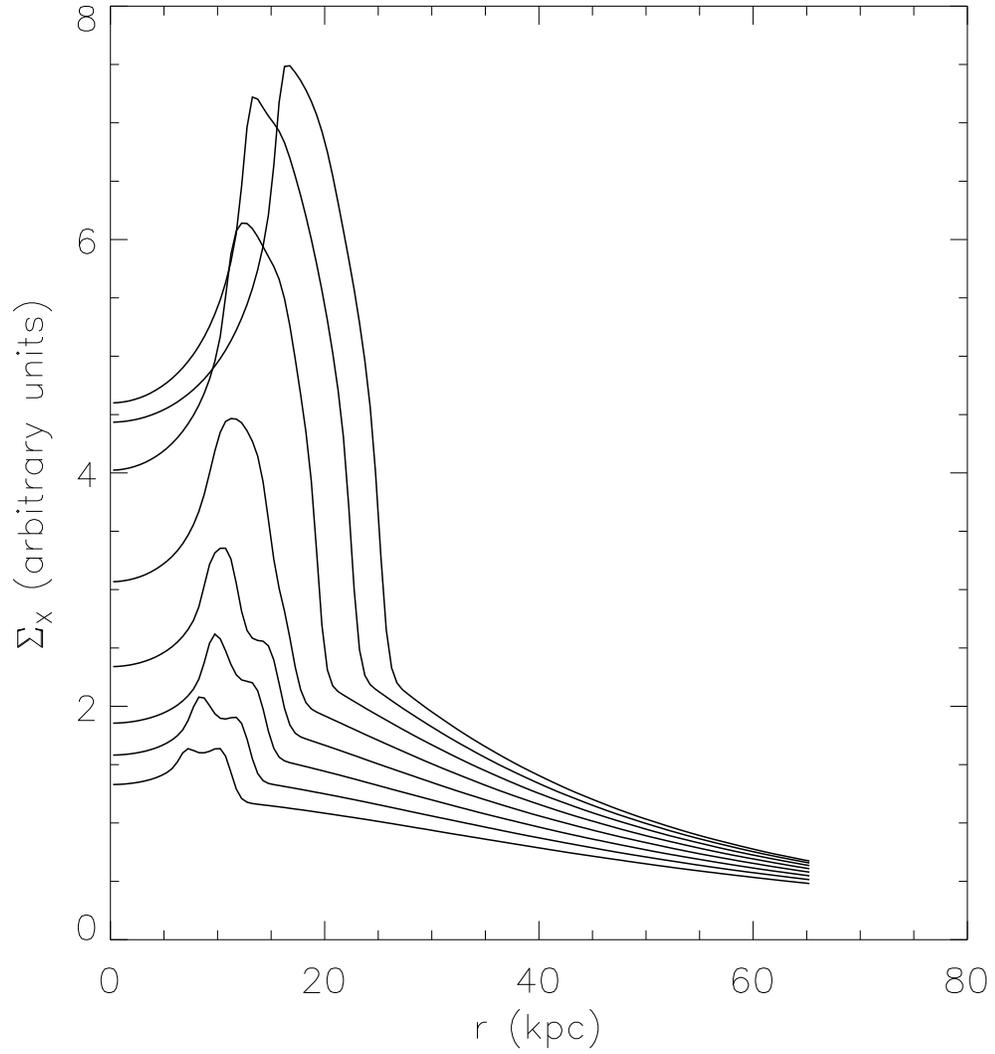}
\vskip0.5in
\caption{
Eight scans of the bolometric X-ray surface brightess 
$\Sigma_X$ perpendicular to the jet direction for model 1 
(bottom panel of Fig. 3). 
From top to bottom the scans are at 
$z = 10$, 15, 20, 25, 30, 35, 40, and 45 kpc from the 
center of Cygnus A. 
}
\label{f12}
\end{figure}

\end{document}